\documentclass[12pt]{article}
\usepackage{a4wide,amssymb,cite}
\pdfoutput=1

\usepackage{a4wide,amssymb,graphicx}
\usepackage{epsfig}
\usepackage[usenames,dvipsnames]{color}
\usepackage{slashed}
\usepackage{amssymb,cite,graphicx}
\usepackage{slashed}
\usepackage{amsmath,bm,bbm}
\usepackage{amsfonts}
\usepackage[titletoc,title]{appendix}
\usepackage[small]{caption}
\usepackage[margin=1in]{geometry}
\usepackage[multiple]{footmisc}
\usepackage{mathtools}
\usepackage{slashed}
\usepackage[nottoc]{tocbibind}
\usepackage{xcolor}

\newcommand{\be}{\begin{equation}}
\newcommand{\ee}{\end{equation}}
\newcommand{\bea}{\begin{eqnarray}}
\newcommand{\eea}{\end{eqnarray}}

\def\circa#1{\,\raise.3ex\hbox{$#1$\kern-.75em\lower1ex\hbox{$\sim$}}\,}

\begin{document}

\begin{titlepage}

%
%


%

\begin{centering}
\vspace{1cm}
{\Large {\bf Lectures on Physics Beyond the Standard Model} } \\

\vspace{1.5cm}

{\bf  
 Hyun Min Lee }
\\
\vspace{.5cm}

{\it Department of Physics, Chung-Ang University, Seoul 06974, Korea.} 
 \\ \vspace{0.2cm}

(Email: hminlee@cau.ac.kr)

\end{centering}
\vspace{2cm}

\begin{abstract}
\noindent
We give a brief overview on the successes and theoretical problems of the  Standard Model and discuss  TeV-scale supersymmetry and some of recent  proposals for physics beyond the Standard Model and dark matter physics.

\end{abstract}

\vspace{6.5cm}

\vspace{2cm}

\end{titlepage}

\section{Introduction}

We have been determining all the parameters of the Standard Model (SM) for particle physics with more precision for decades and exploring for signs for new physics that would provide answers to the fundamental origins of the inner structure. Searches for new physics at the Large Hadron Collider have been mainly motivated by the solutions for the Higgs mass hierarchy problem, such as low-scale supersymmetry, composite Higgs models, extra dimensions, etc. 
Furthermore, the SM is never complete for many other reasons, among which dark matter and dark energy issues are most compelling.  

Low-scale supersymmetry has given us a unified picture for nature, stabilizing the Higgs mass as it should be, and unifying the gauge couplings as well as providing a dark matter candidate, etc.  The mass of the discovered Higgs boson and the null results in searches for supersymmetric particles, however, have cast doubts on the realization of low-scale supersymmetry that we had expected. Nonetheless, supersymmetry is still an important guideline for physics beyond the SM in the next decades and it can be considered to be a UV completion in one form or another of many of new ideas proposed for solving the little hierarchy problem. 

In these lectures, we begin with reviewing the basic structure of the SM and making a brief diagnosis of some of the theoretical problems. The basic concept of supersymmetry is given and it is applied to the Minimal Supersymmetric Standard Model (MSSM). Some of pros and cons of the supersymmetric model were addressed and some solutions were given. Then, we continue to introduce alternative ideas for solving the hierarchy problem, such as extra dimensions, clockwork mechanism, relaxion models, twin Higgs models, four-form flux models, etc. Finally,  we also present a discussion on the production mechanism for the Weakly Interacting Massive Particles (WIMP), and make a detailed comparison to self-interacting dark matter with more emphasis on Strongly Interacting  Massive Particles (SIMP) and forbidden dark matter.

\section{The Standard Model}

We give a brief summary of the Standard Model (SM) and the flavor structure after electroweak symmetry breaking.  Focus is given on the presence of nicely protected global symmetries in the SM, such as $B$ and $L$ numbers. Some reviews on the SM can be found in Ref.~\cite{sm1,sm2,sm3}.

\subsection{The Standard Model Lagrangian}

The Standard Model has $SU(3)_C\times SU(2)_L\times U(1)_Y$ gauge symmetries, with corresponding gauge fields, $g^a_\mu$ with $a=1,\cdots, 8$, $W^i_\mu$ with $i=1,2,3$ and $B_\mu$. 
The covariant derivative is
\bea
D_\mu\psi= \Big(\partial_\mu -\frac{1}{2} i g_S \lambda^a G^a_\mu - \frac{1}{2} i  g\tau^i W^i_\mu - i  g' Y B_\mu\Big)\psi
\eea
where for a fundamental representation $\psi$ of $SU(3)_C\times SU(2)_L$, $\lambda^a$ are Gell-Mann matrices and $\tau^i$ are Pauli matrices. We note that $[\frac{\lambda^a}{2},\frac{\lambda^b}{2}]=if^{abc} \frac{\lambda^c}{2} $, $[\frac{\tau^i}{2},\frac{\lambda^i}{2}]=i\epsilon^{ijk} \frac{\tau^k}{2}$, ${\rm Tr}(\lambda^a \lambda^b)=2 \delta^{ab}$ and ${\rm Tr}(\tau^i \tau^j)=2 \delta^{ij}$.

Moreover, there are three copies of quarks and leptons and one Higgs doublet, in the following representations under the SM gauge symmetries:
\bea
q=\left(\begin{array}{cc} u_L \\ d_L \end{array} \right)=(3,2)_{+\frac{1}{6}}, \quad u_R=(3,1)_{+\frac{2}{3}},\quad d_R = (3,1)_{-\frac{1}{3}},
\eea
\bea
l=\left(\begin{array}{cc} \nu_L \\ e_L \end{array} \right)=(1,2)_{-\frac{1}{2}}, \quad e_R=(1,1)_{-1},
\eea
and
\bea
H=\left(\begin{array}{cc} \phi^+ \\ \phi^0 \end{array} \right)=(1,2)_{+\frac{1}{2}}.
\eea
Here, the electromagnetic charge is given by $Q=\tau^3+ Y$. 
Then, the Lagrangian for the SM is
\bea
{\cal L}_{\rm SM}={\cal L}_H +{\cal L}_{\rm G}+ {\cal L}_F +{\cal L}_Y,
\eea
with
\bea
{\cal L}_H&=& |D_\mu H|^2  - m^2_H |H|^2  -\lambda_H |H|^4 , \\
{\cal L}_{ \rm G}&=& -\frac{1}{2} {\rm Tr}(G_{\mu\nu} G^{\mu\nu}) -\frac{1}{2} {\rm Tr}(W_{\mu\nu} W^{\mu\nu}) -\frac{1}{4} B^{\mu\nu} B_{\mu\nu} \\
{\cal L}_F &=& i {\bar q}_L \slashed{D} q_L+i{\bar u}_R \slashed{D} u_R +i {\bar d}_R \slashed{D} d_R + i {\bar l}_L \slashed{D} l_L+i {\bar e}_R \slashed{D} e_R , \\
{\cal L}_Y &=& - y_d {\bar q}_L d_R H-y_u {\bar q}_L u_R {\tilde H}-y_e {\bar l}_L e_R H  +{\rm h.c.}\label{yukawas}
\eea
where $\slashed{D} q_L=\gamma^\mu D_\mu q_L$, etc,  and  ${\tilde H}=i\tau^2 H^*$. 
Here, the field strength tensors are $G_{\mu\nu}=\partial_\mu G_\nu -\partial_\nu G_\mu +ig_S [G_\mu,G_\nu]$ for $G_\mu=\frac{1}{2}\lambda^a G^a_\mu$,  $W_{\mu\nu}=\partial_\mu W_\nu -\partial_\nu W_\mu +ig [W_\mu,W_\nu]$ for $W_\mu=\frac{1}{2}\tau^i W^i_\mu$ and $B_{\mu\nu}=\partial_\mu B_\nu -\partial_\nu B_\mu $. We note that the $SU(2)_L$ invariants are expanded such as ${\bar q}_L H={\bar u}_L \phi^++ {\bar d}_L \phi^0$, etc. 

From the minimization of the Higgs potential, the VEV of the Higgs field is given by
\bea
v=\sqrt{2}\langle |H| \rangle = \sqrt{-\frac{m^2_H}{\lambda_H}}.
\eea
Then, $W$ and $Z$ bosons receive masses,
\bea
m^2_W = \frac{1}{4} g^2 v^2,\quad m^2_Z = \frac{1}{4} (g^2+g^{\prime 2}) v^2.
\eea
Here, the Higgs VEV is determined to be $v=246\,{\rm GeV}$ by the measurement of the Fermi constant with the following relation,
\bea
\frac{G_F}{\sqrt{2}} = \frac{g^2}{8m^2_W} =\frac{1}{2v^2}
\eea
where $G_F=1.16639\times 10^{-5}\,{\rm GeV}^{-2}$ is  
given by the muon decay, $\mu\rightarrow e {\bar\nu}_e \nu_\mu$. 
On the other hand, for $H=\frac{1}{\sqrt{2}}(0,v+h)^T$, the Higgs boson mass determines the Higgs mass parameter by
\bea
m_h=\sqrt{2\lambda_H} \, v = \sqrt{2}|m_H|=125\,{\rm GeV}. 
\eea

\subsection{Flavor structure}

After electroweak symmetry breaking, the Yukawa couplings for fermions determine mass matrices for quarks and leptons,
\bea
{\cal L}_{\rm mass}=- {\bar u}_{iL} m_{u,ij} u_{jR} - {\bar d}_{iL} m_{d,ij} d_{jR} - {\bar e}_{iL} m_{e,ij} e_{jR} +{\rm h.c.}
\eea
with 
\bea
m_{u,ij} = \frac{1}{\sqrt{2}} y_{u,ij}\,v, \quad m_{d,ij} = \frac{1}{\sqrt{2}} y_{d,ij}\,v,\quad  m_{e,ij}= \frac{1}{\sqrt{2}} y_{e,ij}\,v.
\eea
In this case, the original flavor symmetries in the quark sector are broken to global baryon symmetry $U(1)_B$ as
\bea
U(3)_q\times U(3)_u \times U(3)_d \rightarrow U(1)_B.
\eea
As a result, there are 26 broken generators, leaving 9 physical parameters among $18+18$ in $y_{u} $ and $y_d$: 6 quark masses and 3 mixing angles and one CP phase.

We can diagonalize the fermion mass matrices by bi-unitary transformations to
\bea
m^{\rm diag}_u=V_{u L} m_u V^\dagger_{u R}, \quad m^{\rm diag}_d=V_{d L} m_u V^\dagger_{d R}, \quad m^{\rm diag}_e=V_{e L} m_u V^\dagger_{e R},
\eea
with mass eigenstates being
\bea
f'_{Li} =(V_{fL})_{ij} f_{Lj}, \quad f'_{Ri} = (V_{fR})_{ij} f_{Rj}.
\eea
As a result, the charged current weak interactions for quarks become
\bea
\frac{g}{2} \,{\bar q}_{i} \gamma^\mu (W^1_\mu \tau^1+W^2_\mu \tau^2) q_{i} = \frac{g}{\sqrt{2}} ({\bar u}'_L, {\bar c}'_L, {\bar t}'_L) \gamma^\mu W^+_\mu  V_{\rm CKM} \left( \begin{array}{c} d'_L \\ s'_L  \\ b'_L\end{array} \right)+{\rm h.c.}
\eea
where $W^\pm_\mu=(W^1_\mu \mp i W^2_\mu)/\sqrt{2}$ and  $V_{\rm CKM}$ is the Cabibbo-Kobayashi-Maskawa (CKM) matrix, given by
\bea
V_{\rm CKM} = V_{u L} V^\dagger_{d L}. 
\eea
Thus, charged current weak interactions are the only flavor changing interactions in the SM. 

On the other hand, the neutral current interactions for quarks come from both $SU(2)_L$ and $U(1)_Y$ couplings as 
$$
{\bar q}_{i} \gamma^\mu \Big(\frac{g}{2} \tau^3 W^3_\mu+g' Y_q B_\mu\Big) q_{i} + g' {\bar u}_{Ri}\gamma^\mu Y_u B_\mu u_{Ri} + g' {\bar d}_{Ri}\gamma^\mu Y_d B_\mu d_{Ri} 
$$
\bea
&=& \frac{g}{2\cos\theta_W}\, Z_\mu \Big({\bar u}_i \gamma^\mu (v_u -a_u \gamma_5) u_i + {\bar d}_i \gamma^\mu (v_d -a_d \gamma_5) d_i\Big) +e Q_u A_\mu {\bar u}_i \gamma^\mu u_i  +e Q_d A_\mu {\bar d}_i \gamma^\mu d_i  \nonumber \\
&=&  \frac{g}{2\cos\theta_W}\, Z_\mu \Big({\bar u}'_i \gamma^\mu (v_u -a_u \gamma_5) u'_i + {\bar d}'_i \gamma^\mu (v_d -a_d \gamma_5) d'_i\Big) +e Q_u A_\mu {\bar u}'_i \gamma^\mu u'_i  +e Q_d A_\mu {\bar d}'_i \gamma^\mu d'_i  \nonumber \\
\eea
with $v_u=\frac{1}{2}(1-4 |Q_u| \sin^2\theta_W) $, $v_d=-\frac{1}{2}(1-4 |Q_d| \sin^2\theta_W) $, $a_u=-a_d=\frac{1}{2}$.  Here, the Weinberg mixing angle is introduced as $W^3_\mu =Z_\mu\cos\theta_W  + A_\mu \sin\theta_W$ and $B_\mu=-Z_\mu \sin\theta_W + A_\mu \cos\theta_W$, and the electroweak gauge couplings are related to the electromagnetic coupling by  $e=g\sin\theta_W=g'\cos\theta_W$.  
Therefore, there is no Flavor Changing Neutral Currents (FCNC) for quarks at tree level in the SM, thanks to the Glashow-Iliopoulos-Maiani (GIM) mechanism. 

Similarly, the charged current weak interactions for leptons are
\bea
\frac{g}{2} \,{\bar l}_i \gamma^\mu (W^1_\mu \tau^1+W^2_\mu \tau^2)  l_i = \frac{g}{\sqrt{2}} ({\bar \nu}'_{e L}, {\bar \nu}'_{\mu L}, {\bar \nu}'_{\tau L}) \gamma^\mu W^+_\mu  U_{\rm PMNS} \left( \begin{array}{c} e'_L \\ \mu'_L  \\ \tau'_L\end{array} \right)+{\rm h.c.}
\eea
where $U_{\rm PMNS}$ is the Pontecorvo-Maki-Nakagawa-Sakata (PMNS) matrix, given by
\bea
U_{\rm PMNS}= V_{\nu L} V^\dagger_{e L}.
\eea
When the charged lepton matrix is diagonal, $U_{\rm PMNS}=V_{\nu L}$ is just the mixing matrix for neutrinos. The neutral current interactions for leptons are similarly given by
$$
{\bar l}_{i} \gamma^\mu \Big(\frac{g}{2} \tau^3 W^3_\mu+g' Y_l B_\mu\Big) l_{i} + g' {\bar e}_{Ri}\gamma^\mu Y_e B_\mu e_{Ri} 
$$
\bea
&=& \frac{g}{2\cos\theta_W}\, Z_\mu \Big({\bar e}_i \gamma^\mu (v_e -a_e \gamma_5) e_i + {\bar \nu}_i \gamma^\mu (v_\nu -a_\nu \gamma_5) \nu_i\Big) -e A_\mu {\bar e}_i \gamma^\mu e_i  \nonumber \\
&=& \frac{g}{2\cos\theta_W}\, Z_\mu \Big({\bar e}'_i \gamma^\mu (v_e -a_e \gamma_5) e'_i + {\bar \nu}'_i \gamma^\mu (v_\nu -a_\nu \gamma_5) \nu'_i\Big) -e A_\mu {\bar e}'_i \gamma^\mu e'_i  
\eea
where $v_e=\frac{1}{2}(-1+4\sin^2\theta_W) $, $a_e=-\frac{1}{2}$, and  $v_\nu=a_\nu=\frac{1}{2}$.
Again there is no FCNC for leptons at tree level. 

The Standard Model has global symmetries respected at the level of dimension-4 operators, that are $U(1)_B$ and $U(1)_L$ associated with baryon and lepton numbers, respectively.  $B=\frac{1}{3}$ is assigned for quarks and $L=1$ for leptons. 
There are no masses for neutrinos in the SM, so it is necessary to extend the SM with new interactions and/or particle content.

\section{Problems in the Standard Model}

We discuss the problems and challenges in the SM, focusing on hierarchy problem, vacuum instability problem, flavor problem, gauge coupling unification,  strong CP problem, $L$ and $B$ number violations, and cosmological constant problem. There are other important issues  such as dark matter, baryon asymmetry, inflation, and quantum gravity, etc. Dark matter issue will be touched upon in the later chapter. Reviews and discussions on naturalness problems can be found in Ref.~\cite{naturalness}, and some of recent lectures on physics beyond the Standard Model (BSM) are listed in Ref.~\cite{BSM}.

\subsection{Dimension-2: Hierarchy problem}

The hierarchy problem consists in the huge hierarchy between the Planck mass $M_{Pl}\sim 10^{18}\,{\rm GeV}$ and the Higgs mass parameter $|m_H|\sim 100\,{\rm GeV}$.
This problem becomes manifest in the one-loop correction to the dimension-2 Higgs mass parameter due to top loops, as follows,
\bea
\Delta m^2_H=- \frac{N_c y^2_t}{8\pi^2}\, \Lambda^2 +\cdots
\eea
with $\Lambda$ being the UV cutoff for the loop momentum and typically of order $M_{Pl}$.

Moreover, if a heavy particle  couples to the Higgs doublet, it corrects the Higgs mass parameter a lot. For instance, suppose that a heavy scalar $X$ with mass $M_X$ has a quartic coupling to the Higgs doublet by ${\cal L}_{\rm int}= -\frac{1}{2}\lambda_{HX}X^2 |H|^2$. Then, the one-loop correction to the Higgs mass parameter in dimensional regularization is
\bea
\Delta m^2_H = \frac{\lambda_{HX}}{16\pi^2} \, M^2_X \ln \frac{M^2_X}{\mu^2}. 
\eea
When a right-handed neutrino $\nu_R$ with mass $M_R$ couples to the Higgs doublet by ${\cal L}_{\rm int}=-y_N {\bar l}_L {\tilde H} \nu_R+{\rm h.c.}$, it also contributes to the Higgs mass parameter as
\bea
\Delta m^2_H = \frac{y^2_N}{4\pi^2} \, M^2_R \ln \frac{M^2_R}{\mu^2}. 
\eea
Therefore, unless $\lambda_{HS}$ or $y_N$ is small,  $M$ or $M_R$ much larger than the weak scale leads to a tuning in choosing a correct Higgs mass parameter. 

The simple solutions to the hierarchy problem include low-energy supersymmetry, composite Higgs models (including twin Higgs models), and extra dimensions. Recently, relaxion and clockwork mechanisms are newly proposed too. This issue will be dealth with in the later chapter.

\subsection{Dimension-4: Vacuum instability problem }

The measured Higgs mass infers the Higgs quartic coupling to be $\lambda_H=0.13$ at the electroweak scale. The Yukawa coupling for top quark is determined to $y_t=\sqrt{2}m_t/v$ by the top quark mass. The Higgs quartic couplings runs at high energies, with renormalization group equations at one loop,
\bea
(4\pi)^2 \frac{d\lambda_H}{d\ln\mu}&=& \Big(12y^2_t-3 g^{\prime 2}-9g^2\Big)\lambda_H-6y^4_t + \frac{3}{8}\Big[2g^4 + (g^{\prime 2}+g^2)^2 \Big] +24\lambda^2_H, \\
(4\pi)^2 \frac{d y_t}{d\ln\mu}&=& y_t \Big(\frac{9}{2}y^2_t - 8 g^2_3 -\frac{9}{4} g^2 -\frac{17}{12} g^{\prime 2}\Big).
\eea
Thus, the running quartic coupling at high energy depends on the top Yukawa coupling at low energy. 
We also note that the Higgs mass parameter runs according to
\bea
(4\pi)^2 \frac{d m^2_H}{d\ln\mu} = \Big(12\lambda_H+ 6y^2_t -\frac{9}{2} g^2 - \frac{3}{2} g^{\prime 2} \Big)m^2_H.
\eea

Moreover, the Higgs potential is corrected by one-loop Coleman-Weinberg(CW) in $\overline{\rm MS}$ scheme, given by
\bea
V(h)=\frac{1}{2} m^2_H(\mu) h^2+\frac{1}{4} \lambda_H(\mu) h^4 + V_{\rm CW}(h)
\eea
with 
\bea
V_{\rm CW}(h)= \sum_\alpha \frac{N_\alpha M^4_\alpha}{64\pi^2} \bigg[ \ln\frac{M^2_\alpha}{\mu^2} -C_\alpha\bigg]
\eea
where $\alpha=\{Z,W,t,h,G\}$ for gauge bosons, top quark, Higgs and Goldstones, respectively, with $N_\alpha=\{3,6, -12, 1, 3\}$. Here, Higgs-dependent masses are $M^2_Z=(g^2+g^{\prime 2})h^2/4$, $M^2_W=g^2 h^2/4$, $M^2_t=y^2_t h^2/2$, $M^2_h=3\lambda_H h^2 +m^2_H$, and $M^2_G=\lambda_H h^2 + m^2_H$, and $C_\alpha=\frac{3}{2}$ for fermions or scalars and $C_\alpha=\frac{5}{6}$ for gauge bosons.  For the RG-improved effective potential at higher loops, we need to replace $h$ in the effective potential by the renormalized one, $e^{\Gamma(h)}\,h $, with $\Gamma(h)=\int^h_{m_t} \gamma_h(\mu)d\ln\mu $ where $\gamma_h\equiv d\ln h/d\ln\mu$ is the anomalous dimension of the Higgs field.

For $h\gg v$, we need to take into account the minimization of logarithms in the Coleman-Weinberg potential, because the $n$-loop expansion is valid for $\alpha^{n+1}[\ln (h^2/\mu^2)]^n\leq 1$ with $\alpha={\rm max}(\lambda_H, g^2,g^{\prime 2},y^2_t)/(4\pi)$. Thus, we choose the renormalization scale to $\mu= h$, so the effective potential at $h\gg v$ is dominated by the quartic potential with the effective quartic coupling, as follows,
\bea
V(h)\approx \frac{1}{4} \lambda_{H,{\rm eff}}(h) h^4,
\eea
 with
 \bea
  \lambda_{H,{\rm eff}}(h) = \lambda_H(h)+ \sum_\alpha \frac{N_\alpha M^4_\alpha}{64\pi^2h^4} \bigg[ \ln\frac{M^2_\alpha}{h^2} -C_\alpha\bigg]
 \eea
 where $\lambda_H(h)$ is the running quartic coupling with $\mu$ being replaced by the Higgs field.

Ignoring the CW contributions to the Higgs quartic coupling, the running quartic coupling at $h\gg v$ is given by
\bea
\lambda_H(h)=\lambda_H(v) + \frac{1}{16\pi^2} \bigg[-\frac{12m^4_t}{v^4} + \frac{3}{16} \Big( 2g^4 + (g^{\prime 2}+g^2)^2\Big)+{\cal O}(\lambda_H^2) \bigg] \ln \frac{h^2}{v^2}.
\eea
Then, for $\lambda_H(\Lambda_I)=0$, we obtain 
\bea
\lambda_H(v)= \frac{m^2_h}{2v} =  \frac{1}{16\pi^2} \bigg[\frac{12m^4_t}{v^4} - \frac{3}{16} \Big( 2g^4 + (g^{\prime 2}+g^2)^2\Big)\bigg] \ln \frac{\Lambda^2_I}{v^2}.
\eea
Therefore, the Higgs quartic coupling  turns to a negative value above $\Lambda_I=10^{11}\,{\rm GeV}$ for $m_h=125\,{\rm GeV}$ and $m_t=173\,{\rm GeV}$, so there could be a deep minimum at large Higgs field values, destabilizing the electroweak minimum \cite{vsb1}.  This is called the vacuum instability problem.

A simple solution to the vacuum instability problem is to introduce an extra quartic coupling between the Higgs doublet and a singlet scalar field $S$ \cite{vsb2}, as in the following,
\bea
\Delta V = 2\lambda_{HS} |H|^2 |S|^2+\lambda_S |S|^4. 
\eea
In this case, the extra quartic coupling has two effects for the vacuum stability: one is to modify the renormalization group equations by adding a positive contribution to the beta function of $\lambda_H$ and lifting up the running Higgs quartic coupling to positive values at high energy, as follows,
\bea
(4\pi)^2 \frac{d\lambda_H}{d\ln\mu}&=& \Big(12y^2_t-3 g^{\prime 2}-9g^2\Big)\lambda_H-6y^4_t + \frac{3}{8}\Big[2g^4 + (g^{\prime 2}+g^2)^2 \Big] +24\lambda^2_H+4\lambda^2_{HS}, \\
(4\pi)^2 \frac{d\lambda_{HS}}{d\ln\mu}&=& \frac{1}{2} \Big(12y^2_t -3 g^{\prime 2} -9 g^2  \Big)\lambda_{HS} +4\lambda_{HS} (3\lambda_H +2\lambda_S) + 8\lambda^2_{HS}, \\
(4\pi)^2 \frac{d\lambda_S}{d\ln\mu}&=& 8\lambda^2_{HS} + 20\lambda^2_S. 
\eea
Another more important effect is to make a tree-level shift in the running quartic coupling due to the threshold effect in the presence of the singlet VEV $\langle S\rangle=w/\sqrt{2}$ \cite{vsb2}. After diagonalizing the mass matrix for the singlet and Higgs bosons, 
\bea
{\cal M}^2 = 2\left( \begin{array}{cc}  \lambda_H v^2 & \lambda_{HS} vw \\  \lambda_{HS} vw &  \lambda_S w^2  \end{array} \right),
\eea
the mass eigenvalue of the lightest scalar is given by
\bea
m^2_h= 2v^2 \bigg[\lambda_H -\frac{\lambda^2_{HS}}{\lambda_S} + {\cal O}\Big(\frac{v^2}{w^2}\Big) \bigg].
\eea 
Then, the running Higgs quartic coupling has a shift at tree level by $\lambda_H=0.13+\frac{\lambda^2_{HS}}{\lambda_S} $, at the electroweak scale, so it remains positive after the RG evolution to high energy. 

{\bf Problem}: Check explicitly the scalar threshold corrections to the running quartic coupling.

As will be discussed in the later chapter, in low-energy supersymmetry, the Higgs quartic coupling is given in terms of electroweak gauge couplings and remains positive all the way to the unification scale.  So, there is no vacuum instability problem.

\subsection{Dimension-4: Flavor problem}

There are hierarchies in quark and charged lepton masses in the SM:
\bea 
m^{\rm diag}_u =\left( \begin{array}{ccc} m_u & 0 & 0 \\ 0 & m_c & 0  \\   0 & 0 & m_t \end{array}  \right)= m_t \left( \begin{array}{ccc} 10^{-5} & 0 & 0 \\ 0 & 10^{-3} & 0  \\   0 & 0 & 1 \end{array}  \right) , 
\eea
\bea 
m^{\rm diag}_d =\left( \begin{array}{ccc} m_d & 0 & 0 \\ 0 & m_s & 0  \\   0 & 0 & m_b \end{array}  \right)=m_b \left( \begin{array}{ccc} 10^{-3} & 0 & 0 \\ 0 & 10^{-2} & 0  \\   0 & 0 & 1 \end{array}  \right) , 
\eea
and
\bea 
m^{\rm diag}_e =\left( \begin{array}{ccc} m_e & 0 & 0 \\ 0 & m_\mu & 0  \\   0 & 0 & m_\tau \end{array}  \right)=m_\tau \left( \begin{array}{ccc} 10^{-3} & 0 & 0 \\ 0 & 10^{-1} & 0  \\   0 & 0 & 1 \end{array}  \right).
\eea
Moreover, the CKM mixing matrix are almost diagonal as
\bea
V_{\rm CKM}= \left( \begin{array}{ccc} V_{ud} & V_{us} & V_{ub} \\ V_{cd} & V_{cs} & V_{cb}  \\   V_{td} & V_{ts} & V_{tb} \end{array}  \right)\approx \left(\begin{array}{ccc} 0.974 &  0.225 & 0.0036  \\  0.225  & 0.974  &  0.041  \\  0.009  & 0.040  & 0.999 
\end{array} \right). 
\eea
The CP violation in the quark sector is parametrized by the Jarlskorg invariant $J$ where ${\rm Im}(V_{ij} V_{kl}V^*_{il} V^*_{kj})=J \sum_{m,n} \epsilon_{ikm}\epsilon_{jln}$, and the measured value of $J$ is $J=(3.18\pm 0.15)\times 10^{-5}$.

Furthermore, the neutrino oscillation data determine the differences between neutrino masses and mixing matrix at best fit, as follows \cite{pdg},
\bea
|\Delta m^2_{21}| &=& 7.37\times 10^{-5}\, {\rm eV}^2, \\
|\Delta m^2_{23}| &=& 2.54 \times 10^{-3}\, {\rm eV}^2, 
\eea
and 
\bea
U_{PMNS}&=& \left( \begin{array}{ccc} c_{12} c_{13} & s_{12} c_{13} & s_{13} e^{-i\delta} \\ -s_{12} c_{23} -c_{12} s_{23} s_{13} e^{i\delta} & c_{12} c_{23} -s_{12} s_{23} s_{13} e^{i\delta} & s_{23} c_{13}  \\  s_{12} s_{23} -c_{12} c_{23} s_{13} e^{i\delta} & -c_{12} s_{23} -s_{12} c_{23} s_{13} e^{i\delta} & c_{23} c_{13} \end{array}  \right) \times \nonumber \\
&&\quad\times\, {\rm diag} (1, e^{i\alpha_{21}/2}, e^{i\alpha_{31}/2})
\eea
where  $\sin^2\theta_{12}=0.297$, $\sin^2\theta_{23}=0.425 (0.589)$, $\sin^2\theta_{13}=0.0215(0.0216)$ for $\Delta m^2_{23}<0$: normal hierarchy ($\Delta^2_{23}>0$: inverted hierarchy), and $\delta/\pi=1.38(1.31)$.
The upper bound on the absolute neutrino mass is obtained from the spectrum of electrons near the end point in the ${}^3H$ $\beta$-decay experiments. The Troitzk experiment set the limit $m_{{\bar \nu}_e}<2.05\,{\rm eV}$ at $95\%$ CL, and the upcoming KATRIN can reach sensitivity of $m_{{\bar \nu}_e}\sim 0.20\,{\rm eV}$.  The CMB data as well as BAO set the upper limit on the sum of neutrino masses, $\sum_j m_{\nu_j}<0.170\,{\rm eV}$ at $95\%$ CL.

The flavor problem is about the hierarchical patterns of fermion masses and the mixing patterns of quarks and leptons.  In particular, neutrino masses are much lighter than any of quarks and leptons, $m_{\nu_j}/m_{l,q}\lesssim 10^{-6}$. The flavor problem is originated from the dimension-4 Yukawa couplings for quarks and leptons, and partly from the dimension-5 operators for neutrino masses.

\subsection{Dimension-4: Gauge coupling unification}

There are three independent gauge couplings, $g_S, g, g'$, for strong, weak and hypercharge (or electromagnetic) interactions in the SM, respectively.
The measured values of the gauge couplings at low energy are $g(m_t)=0.64$, $g'(m_t)=0.35$ and $g_S(m_t)=1.16$.  The gauge couplings in the SM run in energy by the RG equations,
\bea
\frac{d\alpha^{-1}_i}{d\ln\mu} = - \frac{b_i}{2\pi}, \qquad i=1,2,3,
\eea
where $\alpha_i=\frac{g^2_i}{4\pi}$ with $g_1=\sqrt{\frac{5}{3}}\, g'$, $g_2=g$ and $g_3=g_S$, and $b_i$ are the corresponding beta function coefficients,  $b_i= (\frac{41}{10},-\frac{19}{6},-7)$.
But, the gauge couplings do not quite unify even at high scales, so we need to extend the SM for gauge coupling unification. 

{\bf Problem}: The definition for the beta function coefficient is
\bea
b= -\frac{11}{3} C_2(G) +\frac{1}{3}\, C(r_S) + \frac{2}{3} C(r_F) 
\eea
where the quadratic Casimirs are $C_2(G)=N$ for $SU(N)$ and $C_2(G)=0$ for $U(1)$, and $C(r) $ is the Dynkin index for the representation $r$ (for instance, $C(r)=\frac{1}{2} (N)$ for a fundamental(adjoint) representation of $SU(N)$), and $r_S (r_F)$ stands for the representation of a complex scalar (Weyl fermion). Then, evaluate the beta function coefficients for gauge couplings in the SM.

\subsection{Dimension-4: Strong CP problem}

The QCD Lagrangian has an additional gauge-invariant term, the so called $\theta$ term,
\bea
{\cal L}_{\theta}=  \theta\,\frac{g^2}{32\pi^2}\, G^a_{\mu\nu} {\tilde G}^{\mu\nu}_a
\eea
with ${\tilde G}^{\mu\nu}_a=\frac{1}{2}\epsilon^{\mu\mu\rho\sigma} G^a_{\rho\sigma}$. 
It turns out that the $\theta$ term is a total derivative,
\bea
\frac{g^2}{32\pi^2}\, G^a_{\mu\nu} {\tilde G}^{\mu\nu}_a= \partial_\mu K^\mu
\eea
with 
\bea
K^\mu =\frac{g^2}{32\pi^2}\,  \epsilon^{\mu\nu\rho\sigma} G^a_\nu \Big[G^a_{\rho\sigma} -\frac{g}{3} f^{abc}G^b_\rho  G^c_\sigma \Big].
\eea
Therefore, the $\theta$ term does not affect the local QFT properties, but there is a vacuum gauge configuration with a nontrivial topological (winding) number, $n\neq 0$, due to
\bea
\frac{g^2}{32\pi^2} \int d^4 x G^a_{\mu\nu} {\tilde G}^{\mu\nu}_a= \int dS^\mu K_\mu=n
\eea
with $n$ being integer. The non-perturbative effects are proportional to $e^{-c/g^2}$ so only the QCD $\theta$ term is important. 
Indeed, the QCD $\theta$ term contributes to neutron electric dipole moment (EDM) as
\bea
d_n=\frac{e}{\Lambda^2_{\rm QCD}}\, \frac{m_u m_d}{m_u+m_d}\,\theta < 3.0\times 10^{-26}\, e\,{\rm cm},
\eea
which sets the limit to $|\theta|<10^{-10}$.  This is the strong CP problem.
The axion is a dynamical solution to the strong CP problem \cite{axion1,axion2,axionlecture}.

{\bf Problem}: Show that $F_{\mu\nu} {\tilde F}^{\mu\nu}$ for a $U(1)$ gauge theory is a total derivative. Also show that the QCD $\theta$ term is a total derivative.

\subsection{Dimension-5: Lepton number violation}

In the SM, neutrinos are massless at the level of dimension-4 operators, so we need to introduce the Weinberg operator for neutrino masses at dimension-5 level,
\bea
{\cal L}_{\rm dim-5}=- \frac{c_{ij}}{M} (\overline{l^c_i} i\tau^2 H)(l_j i\tau^2 H)+{\rm h.c.}
\eea
where $c_{ij}$ is a $3\times 3$ antisymmetric complex matrix.
Then, the above dimension-5 operator violates the lepton number by two units.
After electroweak symmetry breaking, the above interactions lead to Majorana neutrino mass terms, ${\cal L}_\nu=-m_\nu \overline{\nu^c} \nu +{\rm h.c.}$ with $m_{\nu,ij}=\frac{c_{ij} v^2}{M}$, which is less than $0.1\,{\rm eV}$, resulting in the following lower limit,
$M/|c_{ij}|>10^{14}\,{\rm GeV}$.
Therefore, for $|c_{ij}|={\cal O}(1)$, new physics scale is of order $10^{14}\,{\rm GeV}$.

In the presence of charged lepton Yukawa couplings and Majorana neutrino masses, the original flavor symmetries $U(3)_l\times U(3)_e$ in the lepton sector are broken completely. So, there are 18 broken generators, leaving 12 physical parameters among $18+12$ in $y_e$ and $c$: 6 masses, 3 mixing angles and 3 CP phases.

\subsection{Dimension-6: Baryon number violation}

The dimension-6 operators for $B/L$ violations in the SM are 
\bea
{\cal L}_{\rm dim-6}=\frac{1}{M^2} (y^2_1 qqql + y^2_2 u^c e^c u^c d^c) + \frac{g^2}{M^2} (\overline{d^c}\, \overline{u^c} ql -\overline{e^c} \, \overline{u^c} qq ).
\eea
From the proton lifetime, 
\bea
\tau(p\rightarrow e^+ \pi^0)\sim \frac{M^4}{g^4 m^5_p}>1.6\times 10^{34}\,{\rm yrs},
\eea 
we obtain the bound, $M/g\gtrsim 10^{16}\,{\rm GeV}$. 
Similarly, from the proton lifetime, 
\bea
\tau(p\rightarrow K^+ \nu)\sim \frac{M^4}{y^4_1 m^5_p}>5.9\times 10^{33}\,{\rm yrs},  
\eea
we obtain the bound, $M/y_1\gtrsim 10^{15}\,{\rm GeV}$. 
As a result, the $B/L$ number conservation should be maintained up to $10^{15}-10^{16}\,{\rm GeV}$, depending on the underlying physics.

\subsection{Dimension-0: Cosmological constant problem}

The Lagrangian for Einstein gravity  contains the cosmological constant $\Lambda$,
\bea
S_E= \int  d^4 x  \sqrt{-g} \Big(\frac{1}{16\pi G} \, R - \Lambda +{\cal L}_{\rm SM}  \Big).
\eea 
The observation constrains the cosmological constant  to be $\Lambda^{1/4}_{\rm obs}\sim 10^{-3}\,{\rm eV}$ while the theoretical prediction for $\Lambda$ in the SM is $\Lambda^{1/4}_{\rm th}\sim M_{Pl}$. So, the disparity associated with 
\bea
\frac{\Lambda_{\rm obs}}{\Lambda_{\rm th}}\sim 10^{-120},
\eea 
is the cosmological constant problem. Furthermore, all the unrelated components such as zero-point energies due to particles and the potential changes during phase transitions  (QCD and electroweak phase transitions) contribute to the cosmological constant, which is somehow designed to vanish.

\section{Supersymmetric Lagrangians and phenomenology}

We review the supersymmetry (SUSY) algebra, the SUSY spectrum, and SUSY transformations, and discuss the construction of supersymmetric theories in terms of superfields in superspace. We also review mediation mechanisms for SUSY breaking and the phenomenology in the minimal extension of the SM with SUSY.
Discussion on SUSY and notations are based on Wess and Bagger \cite{WB}.

\subsection{SUSY algebra}

Supersymmetry is a symmetry between bosons and fermions, the maximal extension of spacetime symmetry. It might be manifest at high energy beyond the Standard Model (SM) or emergent at low energy as new phases of matter or optical properties.  Supersymmetry is endowed by string theory as the best candidate for quantum theory of gravity and it is a solution to the hierarchy problem in the SM.

\subsubsection{Coleman-Mandula theorem}

The assumptions for no-go theorem of Coleman and Mandula are: \\
 1. Existence of S-matrix at almost all energies in local relativistic QFT in 4D. \\
2. The number of different particles associated with one-particle states of a given mass is finite. \\
 3. There is an energy gap between vacuum and one-particle states. \\
$\Rightarrow$ The most general Lie algebra of symmetry operators that commute with S-matrix consists of the generators $P_m$ and $M_{mn}$ of the Poincar\'e group, and ordinary internal symmetry generators $B_l$. The latter act on one-particle states with matrices that are diagonal and {\it independent of both momentum and spin}. Here, there are only possible bosonic conserved quantities.

\subsubsection{Graded Lie algebra}

SUSY algebra  (or superalgebra) is the only graded Lie algebra of symmetries of the S-matrix consistent with relativistic QFT.  It extends the Poincar\'e group by anti-commutators, given in Weyl representation by
\bea
\{Q_\alpha,Q_\beta\}&=&0,\quad \{Q_\alpha,{\bar Q}_{\dot\beta}\}=2(\sigma^m)_{\alpha{\dot\beta}} P_m, \\
{[}Q_\alpha,P_m{]}&=&0, \\
{[} Q_\alpha, M_{mn}{]} &=& \frac{1}{2} (\sigma_{mn}) _\alpha\,^\beta \,Q_\beta
\eea
where $\sigma^m=(1,{\vec \sigma})$ and $\sigma^{mn}=\frac{1}{2}[\sigma^m,\sigma^n]$.
Then, the Hamiltonian is given by
\bea
H=P_0= \frac{1}{4} \sum_\alpha  \{Q_\alpha,Q^\dagger_\alpha\}.
\eea
The consequences of the SUSY algebra are: \\
1. Hamiltonian is non-negative. Ground-state energy vanishes for unbroken SUSY. \\
2. SUSY operator changes the spin of a state by $\frac{1}{2}$. \\
3. $[H,Q_\alpha]=[H,Q^\dagger_\alpha]=0$ implies the same masses for boson and fermion.

There are nice discussion and review articles on the essence of SUSY in the version of quantum mechanics \cite{witten0,susyb}.

\subsection{SUSY spectrum}

1. Massive one-particle states with $P^2=-M^2$. \\
In the rest frame of a massive particle, for which $P_\mu=(-m,0)$, the SUSY algebra becomes
\bea
\{Q_\alpha,{\bar Q}_{\dot\beta}\} &=& -2m\sigma^0_{\alpha{\dot\beta}}=2 m\delta_{\alpha,{\dot\beta}}, \\
\{Q_\alpha, Q_\beta\}&=&\{{\bar Q}_{\dot\alpha} ,{\bar Q}_{\dot\beta}\}=0.
\eea
Then, redefining the SUSY operators as
\bea
a_\alpha\equiv \frac{1}{\sqrt{2m}}\, Q_\alpha, \quad a^\dagger_\alpha=  \frac{1}{\sqrt{2m}}\, {\bar Q}_{\dot\alpha},
\eea
we rewrite the SUSY algebra as the one for a fermionic harmonic oscillator,
\bea
\{a_\alpha, a^\dagger_\beta\}&=&\delta_{\alpha\beta}, \\
\{a_\alpha,a_\beta\}&=&\{a^\dagger_\alpha, a^\dagger_\beta\}=0.
\eea
Suppose that $|\Omega_j\rangle$ is the Clifford vacuum with spin $j$, which is $(2j+1)$-dimensional representation of $SU(2)$, satisfying $a_\alpha|\Omega_j\rangle=0 $. Then, we can construct two excited states from the Clifford vacuum by
\bea
&&\qquad\qquad\qquad a^\dagger_\alpha |\Omega_j\rangle, \\
&&\frac{1}{\sqrt{2}} a^\dagger_\alpha a^\dagger_\beta |\Omega_j\rangle= \frac{1}{2\sqrt{2}} (a^\dagger_\alpha a^\dagger_\beta-a^\dagger_\beta a^\dagger_\alpha)  |\Omega_j\rangle.
\eea
Then, a massive matter multiplet is composed of $(j, j+\frac{1}{2},j-\frac{1}{2},j)$. For instance, for $j=0$, we have two states of spin-0 (or one complex scalar) and one state of spin-$\frac{1}{2}$ (or one Majorana fermion).  For $j=\frac{1}{2}$, we have a massive vector multiplet, composed of one state of spin-0, one state of spin-$\frac{1}{2}$ and one state of spin-1.

2. Massless one-particle states with $P^2=0$. \\
Taking a light-cone coordinate for a massless particle, for which $P_\mu=(-P,0,0,P)$,  the SUSY algebra becomes
\bea
\{Q_\alpha,{\bar Q}_{\dot\beta}\} &=& 2E (-\sigma^0_{\alpha{\dot\beta}}+\sigma^3_{\alpha{\dot\beta}}) = 4E\left(\begin{array}{cc} 1 & 0 \\  0 & 0 \end{array} \right), \\
\{Q_\alpha, Q_\beta\}&=&\{{\bar Q}_{\dot\alpha} ,{\bar Q}_{\dot\beta}\}=0.
\eea
As a result, redefining the SUSY operators as
\bea
a_\alpha&\equiv& \frac{1}{\sqrt{4E}}\, Q_1, \quad a^\dagger_\alpha\equiv \frac{1}{\sqrt{4E}}\, Q_{\dot 1},
\eea
 the SUSY algebra becomes
 \bea
 \{a_1, a^\dagger_1\}&=&1, \\
\{a_2,a^\dagger_2\}=\{a_\alpha,a_\beta\}&=&\{a^\dagger_\alpha, a^\dagger_\beta\}=0.
 \eea
Then, the half of the SUSY generators become totally anti-commuting.
Suppose that $|\Omega_{\lambda}\rangle$ is the lowest helicity state such that $a_1 |\Omega_{\lambda}\rangle=0$. Then, the excited states are constructed by
\bea
a^\dagger_1 |\Omega_\lambda\rangle.
\eea
Then, a massless multiplet is composed of $(\lambda,\lambda+\frac{1}{2})$. 
For instance, for $\lambda=-\frac{1}{2}$, we have states of helicities, $-\frac{1}{2}$ and $0$. In this case, obtaining states of helicities, $0$ and $+\frac{1}{2}$, for $\lambda=0$ and adding them as a CP conjugate, we can make a massless matter multiplet with one state of spin-0 and one state of spin-$\frac{1}{2}$. For $\lambda=-1$ and $\lambda=\frac{1}{2}$, we can also make a massless vector multiplet with one state of spin-1 and one state of spin-$\frac{1}{2}$. Furthermore, for $\lambda=-2$ and $\lambda=\frac{3}{2}$, we have a massless graviton multiplet with one state of spin-2 and one state of  spin-$\frac{3}{2}$ (gravitino).

\subsection{SUSY transformations}

Introduce anti-commuting SUSY transformation parameters, $\xi^\alpha, {\bar\xi}_{\dot\alpha}$, satisfying
\bea
\{\xi^\alpha,\xi^\beta\}=\{\xi^\alpha, Q_\beta\}=\cdots =[P_m,\xi^\alpha]=0. 
\eea
From $\xi Q=\xi^\alpha Q_\alpha$ and ${\bar\xi}{\bar Q}={\bar\xi}_{\dot\alpha} {\bar Q}^{\dot\alpha}$, the SUSY algebra becomes
\bea
{[}\xi Q, {\bar\xi}{\bar Q}{]}&=& 2\xi \sigma^m {\bar \xi} P_m, \\
{[}\xi Q, \xi Q{]}&=& [{\bar\xi}{\bar Q},{\bar\xi}{\bar Q}]=0, \\
{[}P_m, \xi Q{]}&=& [P_m, {\bar\xi}{\bar Q}]=0.
\eea

Infinitesimal SUSY transformations of component fields are
\bea
\delta_\xi A &=& ( \xi Q +{\bar\xi}{\bar Q}) \times A, \\
\delta_\xi \psi &=& ( \xi Q +{\bar\xi}{\bar Q}) \times\psi.
\eea
The SUSY algebra requires  two sequential SUSY transformations to satisfy the closure relation,
\bea
(\delta_\xi \delta_\eta -\delta_\eta\delta_\xi) A&=&2(\eta \sigma^m {\bar\xi}- \xi\sigma^m {\bar\eta}) P_m A \nonumber \\
&=&-2i (\eta \sigma^m {\bar\xi}- \xi\sigma^m {\bar\eta}) \partial_m A,
\eea
which is nothing a total derivative. There is a similar closure relation for $\delta_\xi \psi$. 

Off-shell SUSY contains $A,\psi$ as well as auxiliary field $F$, with the corresponding SUSY transformations,
\bea
\delta_\xi A &=& ( \xi Q +{\bar\xi}{\bar Q}) \times A=\sqrt{2}\xi \psi,  \label{susy1} \\
\delta_\xi \psi &=& ( \xi Q +{\bar\xi}{\bar Q}) \times\psi=i\sqrt{2} \sigma^m {\bar\xi} \partial_m A+ \sqrt{2}\xi F, \label{susy2} \\
\delta_\xi F &=& ( \xi Q +{\bar\xi}{\bar Q}) \times F= i\sqrt{2} {\bar\sigma}^m \partial_m \psi. \label{susy3}
\eea
Then, we can check that the closure relations for the above SUSY transformations are satisfied.
The auxiliary field $F$ is the component field of highest dimension, because it transforms up to a total derivative, not changing into a higher spin state.

{\bf Problem}:  Verify the closure relations for eqs.~(\ref{susy1})-(\ref{susy3}).

The SUSY Lagrangian for a massive supersymmetric multiplet with spin-0 and spin-$\frac{1}{2}$ states is given by
\bea
{\cal L}_{\rm SUSY}= {\cal L}_0 + {\cal L}_m
\eea
with
\bea
{\cal L}_0 &=& i \partial_n {\bar\psi} {\bar\sigma}^n \psi+ A^* \Box A + F^* F, \\
{\cal L}_m &=& m \Big( A F + A^* F^*  -\frac{1}{2} \psi\psi - \frac{1}{2} {\bar\psi}{\bar\psi}\Big). 
\eea
The component field have mass dimensions, $[A]=1$, $[\psi]=\frac{3}{2}$ and $[F]=2$.
We note that the equation for $F$ leads to the mass for $A$, which is the same as the mass for $\psi$, and the fermion kinetic term is equivalent to $-i{\bar\psi}\, {\bar\sigma}^n \partial_n \psi$ up to a total derivative.

\subsection{Superfields}

There is a group element corresponding to the SUSY transformation, 
\bea
G(x,\theta,{\bar\theta}) = e^{i(-x^m P_m + \theta Q + {\bar\theta}{\bar Q} )}.
\eea
Then, using the Hausdorff's formula, $e^A \, e^B=e^{A+B +\frac{1}{2}[A,B]+\cdots}$, the product of SUSY transformations (the left multiplication) is given by
\bea
G(0,\xi,{\bar\xi}) G(x^m,\theta,{\bar\theta})  = G(x^m+i\theta \sigma^m {\bar\xi}-i\xi \sigma^m {\bar\theta},\theta+\xi,{\bar\theta}+{\bar\xi}).
\eea
Therefore, the net effect is a translation in the normal coordinates, as well as translations in the anti-commuting parameters,
\bea
x^m &\rightarrow&  x^m+i\theta \sigma^m {\bar\xi}-i\xi \sigma^m {\bar\theta}, \\
\theta &\rightarrow & \theta+\xi, \\
{\bar\theta} &\rightarrow & {\bar\theta}+{\bar\xi}. 
\eea 
We call $(x^m,\theta,{\bar\theta})$ the superspace, including both the normal and anti-commuting coordinates, so the above coordinate transformation correspond to motion in superspace. 

The corresponding differential operator for the SUSY transformation is $\xi Q+{\bar \xi} {\bar Q}$, with
\bea
Q_\alpha &=& \frac{\partial}{\partial \theta^\alpha} - i\sigma^m_{\alpha{\dot\alpha}} {\bar\theta}^{\dot\alpha} \partial_m, \\
{\bar Q}^{\dot\alpha}&=& \frac{\partial}{\partial {\bar\theta}_{\dot\alpha}} +i\theta^\alpha \sigma^m_{\alpha{\dot\beta}} \epsilon^{{\dot\beta}{\dot\alpha}} \partial_m,
\eea
or 
\bea
{\bar Q}_{\dot\alpha}=-\frac{\partial}{\partial {\bar\theta}^{\dot\alpha}} +i\theta^\alpha \sigma^m_{\alpha{\dot\alpha}} \partial_m.
\eea
Then, we can check explicitly that $\{Q_\alpha,{\bar Q}_{\dot\alpha}\}=2i\sigma^m_{\alpha{\dot\alpha}}\partial_m$ and $\{Q_\alpha,Q_\beta\}=\{{\bar Q}_{\dot\alpha},{\bar Q}_{\dot\beta}\}=0$.
But, as $P_m=-i\partial_m$, the differential operators, $Q_\alpha$ and ${\bar Q}_{\dot\alpha}$ do not satisfy the SUSY algebra. 
Thus, instead we need to take the right multiplication of elements for the SUSY transformation  to get
\bea
G(x^m,\theta,{\bar\theta}) G(0,\xi,{\bar\xi})=G(x^m-i\theta \sigma^m {\bar\xi}+i\xi \sigma^m {\bar\theta},\theta+\xi,{\bar\theta}+{\bar\xi}),
\eea
resulting in SUSY covariant derivatives,
\bea
D_\alpha &=& \frac{\partial}{\partial \theta^\alpha} + i\sigma^m_{\alpha{\dot\alpha}} {\bar\theta}^{\dot\alpha} \partial_m, \\
{\bar D}_{\dot\alpha}&=& -\frac{\partial}{\partial {\bar\theta}^{\dot\alpha}} -i\theta^\alpha \sigma^m_{\alpha{\dot\alpha}} \partial_m,
\eea
which satisfies the SUSY algebra, $\{D_\alpha,{\bar D}_{\dot\alpha}\}=-2i\sigma^m_{\alpha{\dot\alpha}}\partial_m$ and $\{D_\alpha,D_\beta\}=\{{\bar D}_{\dot\alpha},{\bar D}_{\dot\beta}\}=0$, and anti-commutators with $Q_\alpha$ and ${\bar Q}_{\dot\alpha}$ vanish.

Superfields $F(x,\theta,{\bar\theta})$ are functions of superspace, $z=(x^m,\theta,{\bar\theta})$. 
Then, $F_1+F_2+\cdots$ and $F_1 F_2\cdots$ are superfields too. 

Chiral superfields $\Phi$ satisfy the constraint, ${\bar D}_{\bar\alpha}\Phi=0$. 
Under the coordinate transformation by
\bea
y^m&=& x^m +i\theta \sigma^m {\bar\theta},  \nonumber \\
\theta'&=&\theta,\qquad {\bar\theta}'={\bar\theta},
\eea
we can show that the SUSY covariant derivatives become
\bea
D_\alpha&=& \frac{\partial}{\partial \theta^{\prime\alpha}} +2 i\sigma^m_{\alpha{\dot\alpha}} {\bar\theta}^{\prime\dot\alpha} \frac{\partial}{\partial y^m},  \label{D1} \\
{\bar D}_{\dot\alpha}&=& -\frac{\partial}{\partial {\bar\theta}^{\prime\dot\alpha}}. \label{D2}
\eea
As a result, the chiral constraint implies that $\Phi=\Phi(y^m, \theta')=\Phi(y^m, \theta)$.

{\bf Problem}: Check eqs.~(\ref{D1}) and (\ref{D2}). 

In turn, the chiral superfields are expanded in the basis of $(x^m, \theta,{\bar\theta})$ as
\bea
\Phi&=&\Phi(y^m,\theta) \nonumber \\
&=& A(y) + \sqrt{2} \theta\psi(y) + \theta\theta F(y) \nonumber \\
&=& A(x) + i\theta \sigma^m {\bar\theta}\, \partial_m A(x) +\frac{1}{4} \theta\theta{\bar\theta}{\bar\theta} \,\Box A(x) \nonumber \\
&&+ \sqrt{2}\theta \psi(x) -\frac{i}{\sqrt{2}} \theta\theta \partial_m\psi \sigma^m {\bar\theta} +\theta\theta F(x)
\eea 
with $y^m=x^m+i\theta\sigma^m {\bar\theta}$.   In this superfield notation, the auxiliary field $F(x)$ appears explicitly as the component field of highest dimension. 
Similarly, an anti-chiral superfield $\Phi^\dagger$ satisfy $D_\alpha\Phi^\dagger=0$, meaning that $\Phi^\dagger=\Phi^\dagger(y^{\dagger m},{\bar\theta})$ with $y^{\dagger m}=x^m-i\theta \sigma^m {\bar\theta}$. Thus, it is also expanded as
\bea
\Phi^\dagger&=& A^*(y^\dagger) + \sqrt{2} {\bar\theta}{\bar\psi}(y^\dagger) + {\bar\theta}{\bar\theta} F^*(y^\dagger) \nonumber  \\
&=& A^*(x) - i\theta \sigma^m {\bar\theta}\, \partial_m A^*(x) +\frac{1}{4} \theta\theta{\bar\theta}{\bar\theta} \,\Box A^*(x) \nonumber \\
&&+ \sqrt{2}{\bar\theta} {\bar\psi}(x) +\frac{i}{\sqrt{2}} {\bar\theta}{\bar\theta}\theta \sigma^m   \partial_m{\bar\psi}(x) +\theta\theta F^*(x).
\eea

Products of chiral superfields are still a chiral superfield. Some products are expanded as
\bea
\Phi_i \Phi_j &=&A_i(y)A_j(y) + \sqrt{2} \theta (\psi_i(y)A_j(y)+ \psi_j(y) A_i(y)) \nonumber \\
&&+\theta\theta \Big( A_i(y) F_j(y)+A_j(y) F_i(y)-\psi_i(y) \psi_j(y)\Big),
\eea
\bea
\Phi_i \Phi_j \Phi_j&=&A_i(y)A_j(y)A_k(y) + \sqrt{2} \theta (\psi_i(y)A_j(y)A_k(y)+ \psi_j(y) A_i(y)A_k(y)+\psi_k(y)A_i(y)A_j(y)) \nonumber \\
&&+\theta\theta \Big( A_i(y) F_j(y)A_k(y)+A_j(y) F_i(y)A_k(y)+A_k(y)F_i(y)A_j(k) \nonumber \\
&&-\psi_i(y) \psi_j(y)A_k(y)-\psi_j(y) \psi_k(y)A_i(y)-\psi_k(y) \psi_i(y)A_j(y)\Big).
\eea
Then, as the additional terms from the expansion of $y$ vanish, the coefficients of the $\theta\theta$ terms depend on $x$ and is of the highest dimension, providing the SUSY invariant Lagrangian up to a total derivative.

The product of chiral and anti-chiral superfields is given by
\bea
\Phi^\dagger_i \Phi_j &=&(A^*_i(y^\dagger) + \sqrt{2} {\bar\theta}{\bar\psi}_i(y^\dagger) + {\bar\theta}{\bar\theta} F^*(y^\dagger)_i )(A_j(y) + \sqrt{2} \theta\psi_j(y) + \theta\theta F_j(y) ) \nonumber \\
&=&\cdots +\theta\theta{\bar\theta}{\bar\theta} \bigg[-\frac{1}{2}\partial^m A^*_i \partial_m A_j+ \frac{1}{4} A_j \Box A_i^* +\frac{1}{4} A^*_i\Box A_j  \nonumber \\
&&\quad+\frac{1}{2}i \partial_n\psi_j \sigma^n {\bar\psi}_i - \frac{1}{2} i\psi_j \sigma^n\partial_n {\bar\psi}_i +F^*_i F_j\bigg].
\eea
Then, the coefficient of the $\theta\theta{\bar\theta}{\bar\theta}$ term, the D-term, is of highest dimension,  becoming SUSY invariant up to a total derivative. 

We keep the coefficients of highest dimension in the above products as the supersymmetric invariant Lagrangian, as they transform as a total derivative only under SUSY transformations. 
For instance, the polynomials of chiral superfields are still superfields, so the component of highest dimension appears in $\theta\theta$ whereas $\Phi^\dagger_i \Phi_i$ leads to the component of highest dimension in $\theta\theta{\bar\theta}{\bar\theta}$.  Therefore, the most general SUSY renormalizable Lagrangian for chiral superfields is
\bea
{\cal L}_\Phi &=& \Phi^\dagger_i \Phi_i|_{\theta\theta{\bar\theta}{\bar\theta}} + \bigg(\frac{1}{2} m_{ij}\Phi_i \Phi_j +\frac{1}{3} g_{ijk}\Phi_i \Phi_j \Phi_k +\lambda_i \Phi_i \bigg)\bigg|_{\theta\theta}+{\rm h.c.} \nonumber \\
&=& {\cal L}_0(A\rightarrow A_i, \psi\rightarrow \psi_i, F\rightarrow F_i)  \nonumber \\
&&+ m_{ij}\Big(A_i F_j -\frac{1}{2} \psi_i \psi_j\Big) +g_{ijk}(A_i A_j F_k -\psi_i \psi_j A_k)
+\lambda_i F_i +{\rm h.c.} .
\eea
We note that the alternative notations can be used for the SUSY Lagrangian with the integration over the fermionic coordinates, 
\bea
{\cal L}_\Phi = \int d^2\theta d^2{\bar\theta}\, \Phi^\dagger_i \Phi_i + \int d^2\theta \bigg(\frac{1}{2} m_{ij}\Phi_i \Phi_j +\frac{1}{3} g_{ijk}\Phi_i \Phi_j \Phi_k +\lambda_i \Phi_i \bigg)+{\rm h.c.}. 
\eea

The Euler equations for auxiliary fields $F_i$ are
\bea
\frac{\partial {\cal L}_\Phi}{\partial F_k} = F^*_k + \lambda_k + m_{ik} A_i + g_{ijk} A_i A_j =0.
\eea
Then, plugging the solution for $F_k$ in the Lagrangian, we obtain the F-term potential,
\bea
V_F &=&-F^*_k F_k -(\lambda_k + m_{ik} A_i + g_{ijk} A_i A_j) F_k +{\rm h.c.} \nonumber \\
&=& F^*_k F_k \nonumber \\
&=& |\lambda_k + m_{ik} A_i + g_{ijk} A_i A_j|^2\geq 0.
\eea
Introducing the superpotential $W$ as a holomorphic function of $\Phi_i$,
\bea
W(\Phi)=\frac{1}{2} m_{ij}\Phi_i \Phi_j +\frac{1}{3} g_{ijk}\Phi_i \Phi_j \Phi_k +\lambda_i \Phi_i,
\eea
we can obtain the Lagrangian for chiral superfields in terms of the F-term potential and the fermion bilinear terms as
\bea
{\cal L}_\Phi = {\cal L}_0+ {\cal L}_{\psi^2}-V_F 
\eea
with
\bea
V_F &=& \sum_i \Big|\frac{\partial W}{\partial \Phi_i} \Big|^2, \\
{\cal L}_{\psi^2}&=& -\frac{1}{2} \sum_{i,j} \frac{\partial^2 W}{\partial \Phi_i \Phi_j} \psi_i\psi_j+{\rm h.c.}
\eea
where the derivatives of the superpotential are evaluated at $\Phi_i=A_i$.

Vector superfields satisfy the reality condition $V=V^\dagger$. 
In Wess-Zumino gauge, vector superfields contain only $v_m$,  $\lambda$ and auxiliary field $D$, written as
\bea
V= -\theta \sigma^m {\bar\theta} v_m(x) + i\theta\theta {\bar\theta} {\bar\lambda}(x) -i {\bar\theta }{\bar\theta} \theta\lambda(x) +\frac{1}{2} \theta\theta {\bar\theta}{\bar\theta} \, D(x).
\eea
We can check that 
\bea
V^2= -\frac{1}{2}  \theta\theta {\bar\theta}{\bar\theta}\, v_m v^m,
 \eea
 and $V^3=0$.  
 
 {\bf Problem}: Verify the above results. 
 
Similarly for chiral superfields, the SUSY transformations for component fields in a vector superfield are
\bea
\delta_\xi v_{m} &=& ( \xi Q +{\bar\xi}{\bar Q}) \times v_m=i\xi \sigma^m {\bar\lambda}+i{\bar\xi}{\bar\sigma}^m \lambda, \\
\delta_\xi \lambda &=& ( \xi Q +{\bar\xi}{\bar Q}) \times\lambda= \sigma^{mn} {\xi} v_{mn}+ i \xi D. \\
\delta_\xi D &=& ( \xi Q +{\bar\xi}{\bar Q}) \times D=  {\bar\xi}{\bar\sigma}^m \partial_m \lambda -\xi\sigma^m \partial_m {\bar\lambda}.
\eea
Therefore, the  auxiliary field $D$ is the component of highest dimension, transforming as a total derivative under SUSY transformation.
 
 The superfield strength is
 \bea
 W_\alpha &=& -\frac{1}{4} {\bar D}{\bar D} D_\alpha V \nonumber \\
 &=&-i\lambda_\alpha(y) +\theta_\beta \Big[\delta^\beta_\alpha D(y)-\frac{i}{2} (\sigma^n {\bar\sigma}^m)_\alpha\,^\beta (\partial_n v_m(y)-\partial_m v_n(y)) \Big]  \nonumber \\
 &&+\theta\theta \sigma^m_{\alpha{\dot\alpha}} \partial_m {\bar\lambda}^{\dot\alpha}(y),
 \eea
 The superfield strength is a chiral superfield satisfying ${\bar D}_{\dot\beta} W_\alpha=0$ and it contains only gauge invariant fields, $\lambda_\alpha, D$ and $\partial_n v_m -\partial_m v_n$. 
 Then,  from 
 \bea
 W^\alpha W_\alpha|_{\theta\theta} = -2i\lambda \sigma^m \partial_m {\bar\lambda}+ D^2  -\frac{1}{2} f_{nm} f^{nm} -\frac{1}{4} i \epsilon^{klnm} f_{kl} f_{nm},
 \eea
we obtain the kinetic terms for component fields in vector superfields as
\bea
{\cal L}_V &=&  \frac{1}{4} \Big(  W^\alpha W_\alpha|_{\theta\theta}+{\rm h.c.}\Big) \nonumber \\
&=& -i\lambda  \sigma^m \partial_m {\bar\lambda} + \frac{1}{2} D^2 -\frac{1}{4} f_{nm} f^{nm} 
\eea
with $f_{nm}=\partial_n v_m -\partial_m v_n$.
The above SUSY Lagrangian can be rewritten by the integration over the fermionic coordinates, 
\bea
{\cal L}_V = -\frac{1}{4}\int d^2\theta \, W^\alpha W_\alpha + {\rm h.c.}. 
\eea

\subsection{Supersymmetric gauge theories}

When chiral superfields $\Phi_\pm$ carry opposite charges under local $U(1)$, they transform under the gauge transformation as
\bea
\Phi_\pm\rightarrow e^{\mp 2 ie \Lambda} \Phi_\pm,
\eea
with $\Lambda$ being a transformation chiral superfield satisfying ${\bar D}_{\dot\alpha}\Lambda=0$, 
while the $U(1)$ vector superfield transforms 
as $V\rightarrow V + \Lambda +\Lambda^\dagger$. For instance, $\Phi_+$ contains electron  $\psi_+$ and $\Phi_-$ contains positron $\psi_-$. Then, $\psi_+, \psi_-$ makes one massive Dirac spinor.

Replacing $\Phi^\dagger_\pm \Phi_\pm$ in the chiral superfield Lagrangian by $\Phi^\dagger_\pm  e^{\pm 2 eV} \Phi_\pm$ and introducing the superpotential, $W=m\Phi_+ \Phi_-$,
the SQED Lagrangian is given by
\bea
{\cal L}_{\rm SQED}&=&\frac{1}{4} \Big(  W^\alpha W_\alpha|_{\theta\theta}+{\rm h.c.}\Big)+ (\Phi^\dagger_+ e^{2eV} \Phi_++\Phi^\dagger_- e^{-2eV} \Phi_-)|_{\theta\theta{\bar\theta}{\bar\theta}} + (W|_{\theta\theta} +{\rm h.c.}) \nonumber \\
&=& -i\lambda  \sigma^m \partial_m {\bar\lambda} + \frac{1}{2} D^2 -\frac{1}{4} f_{nm} f^{nm}  \nonumber \\
&&+ i \partial_m {\bar\psi}_+ {\bar\sigma}^m \psi_+  -m(\psi_+ \psi_- +{\bar\psi}_+ {\bar\psi}_-)  + i \partial_m {\bar\psi}_- {\bar\sigma}^m \psi_-  \nonumber \\
&&+A^*_+ \Box A_+ + A^*_- \Box A_- - V \nonumber \\
&&+ e v^n \bigg[   {\bar\psi}_+ {\bar\sigma}^n \psi_+ + i (A^*_+\partial_n A_+-A_+ \partial_n A^*_+) \nonumber \\
&&\quad \quad +{\bar\psi}_- {\bar\sigma}^n \psi_- +  i (A^*_-\partial_n A_--A_- \partial_n A^*_-) \bigg] \nonumber \\
&&-e^2 v_m v^m (A^*_+ A_+ + A^*_- A_-) \nonumber \\
&&+ i\sqrt{2}e \Big(A^*_+ \psi_+\lambda-A_+{\bar\psi}_+{\bar\lambda}_+-A^*_-\psi_- \lambda+ A_- {\bar\psi}_- {\bar\lambda} \Big) 
\eea
where the total scalar potential 
\bea
V=V_F + V_D \geq 0
\eea
is composed of the F-term potential,
\bea
V_F&=&\Big|\frac{\partial W}{\partial A_+}\Big|^2+\Big|\frac{\partial W}{\partial A_-}\Big|^2 \nonumber \\
&=& m^2 |A_+|^2 + m^2 |A_-|^2,
\eea
and the D-term potential,
\bea
V_D &=& - \frac{1}{2} D^2 - e \, D (A^*_+ A_+ - A^*_- A_-) \nonumber \\
&=& \frac{1}{2} D^2 \nonumber \\
&=&  \frac{e^2}{2} (A^*_+ A_+ - A^*_- A_-)^2.
\eea

Our discussion can be generalized to the case with non-abelian gauge groups. 
In this case, the transformation parameter and vector superfields are generalized to the matrix forms, $\Lambda_{ij}=T^a_{ij} \Lambda_a$ and $V_{ij}=T^a_{ij} V_a$, where $T^a$ is Hermitian and $[T^a,T^b]=i t^{abc} T^c$ and ${\rm Tr}(T^a T^b)=k \delta^{ab}$ with $k>0$ in adjoint representation. Then, the non-abelian gauge transformations are
\bea
\Phi'&=&e^{-2ig\Lambda} \, \Phi, \quad \Phi^{\prime ^\dagger}= \Phi^\dagger e^{2ig\Lambda^\dagger}, \\
e^{2gV'}&=& e^{-2ig\Lambda^\dagger} e^{2gV} e^{2ig\Lambda}. 
\eea
Then, the supersymmetric field strength is also generalized to
\bea
W_\alpha= -\frac{1}{4} {\bar D} {\bar D} \, e^{-2gV} D_\alpha \, e^{2gV}.
\eea
We can check that $W_\alpha$ transforms under the gauge transformation as
\bea
W'_\alpha = e^{-2ig\Lambda} \, W_\alpha\, e^{2ig\Lambda}. 
\eea
Therefore, the general SUSY Lagrangian for non-abelian gauge theories in superspace is
\bea
{\cal L} &=& \frac{1}{4k}\, \bigg[{\rm Tr} (W^\alpha W_\alpha)\Big|_{\theta\theta} +{\rm h.c.}\bigg] + \Phi^\dagger e^{2g V} \Phi \Big|_{\theta\theta {\bar\theta}{\bar\theta}} \nonumber \\
&&+\bigg(\frac{1}{2} m_{ij}\Phi_i \Phi_j + \frac{1}{3} g_{ijk} \Phi_i \Phi_j \Phi_k \bigg)\Big|_{\theta\theta} +{\rm h.c.}. 
\eea
Thus, the component field Lagrangian for gauge interactions only is 
\bea
{\cal L} &=&-i{\bar \lambda}^{(a)}  {\bar \sigma}^m D_m \lambda^{(a)} + \frac{1}{2} D^{(a)} D^{(a)} -\frac{1}{4} f^{(a)}_{nm} f^{(a)nm}  \nonumber \\
&&-i{\bar \psi}\, {\bar \sigma}^m D_m \psi  - D^m A^\dagger D_m A \nonumber \\
&&+ i\sqrt{2}g\Big(A^\dagger T^{(a)} \psi \lambda^{(a)}- {\bar\lambda}^{(a)} T^{(a)}A {\bar\psi} \Big) + g D^{(a)} A^\dagger T^{(a)} A, 
\eea
with
\bea
D_m\psi &=& \partial_m \psi+ ig T^{(a)} v^{(a)}_m \psi, \\
D_m A&=& \partial_m A+ ig T^{(a)} v^{(a)}_m A, \\
 D_m \lambda^{(a)}&=&  \partial_m\lambda^{(a)}  - g\,  t^{abc} v^{(b)}_m \lambda^{(c)}, \\
 f^{(a)}_{mn}&=& \partial_m v^{(a)}_n - \partial_n v^{(a)}_m - g\,  t^{abc} v^{(b)}_m v^{(c)}_n. 
\eea
In this case, the D-term potential is given by
\bea
V_D=\frac{1}{2} D^{(a)}D^{(a)} = \frac{1}{2} g^2 \Big(A^\dagger T^{(a)} A \Big)^2. 
\eea

\subsection{SUSY breaking and mediations}

Supersymmetry is broken in nature due to the null signals for superparticles. 
In this section, we discuss the mechanisms for spontaneous SUSY breaking \cite{susyb} and the transmission of SUSY breaking to the visible sector by messenger interactions.

\subsubsection{Criterion for SUSY breaking}

The fermion number of operator $F$ is defined such that
\bea
(-1)^F =\bigg\{ \begin{array}{c}+1, \quad\,\,\,\, {\rm bosons} \\ -1, \quad {\rm fermions}. \end{array}
\eea
Thus, $(-1)^F Q_\alpha=-Q_\alpha (-1)^F$. As a result, for a finite-dimensional SUSY representation, we take the following trace,
\bea
{\rm Tr} \Big[(-1)^F 2\sigma^m_{\alpha{\dot\beta}} P_m\Big]&=&{\rm Tr} \Big[(-1)^F \{Q_\alpha,{\bar Q}_{\dot\beta} \} \Big]  \nonumber \\
&=& {\rm Tr} \Big[(-1)^F (Q_\alpha{\bar Q}_{\dot\beta}+{\bar Q}_{\dot\beta}Q_\alpha) \Big] \nonumber \\
&=& {\rm Tr} \Big[-Q_\alpha (-1)^F{\bar Q}_{\dot\beta}+{\bar Q}_{\dot\beta} (-1)^FQ_\alpha \Big]=0.
\eea
Then, for fixed $P_m\neq 0$, we get ${\rm Tr}(-1)^F=0$.
Thus, for nonzero energy, the number of bosonic states is the same as the number of fermionic states.

Similarly to the case of quantum mechanics, from the SUSY algebra, we obtain 
\bea
Q_1 {\bar Q}_{\dot 1} + {\bar Q}_{\dot 1} Q_1 &=& 2\sigma^0_{1{\dot 1}} P_0 +2\sigma^i_{1{\bar 1}} P_i, \\
Q_2 {\bar Q}_{\dot 2} + {\bar Q}_{\dot 2} Q_2 &=& 2\sigma^0_{2{\dot 2}} P_0 +2\sigma^i_{2{\bar 2}} P_i.
\eea
Since $\sigma^0_{1{\dot 1}}=\sigma^0_{2{\dot 2}}=-1$, ${\rm tr}(\sigma^i)=0$ and $E=P_0$, adding both equations leads to the Hamiltonian,
\bea
H =\frac{1}{4} (Q_1 {\bar Q}_{\dot 1} + {\bar Q}_{\dot 1} Q_1+Q_2 {\bar Q}_{\dot 2} + {\bar Q}_{\dot 2} Q_2).
\eea
Then,  the Hamiltonian is non-negative. If $\langle 0| H |0\rangle=0$  for the vacuum state $|0\rangle$,  we have $Q_\alpha|0\rangle={\bar Q}_{\bar\alpha}|0\rangle=0$, so $|0\rangle$ is the SUSY-preserving vacuum. 

The Witten index \cite{witten} in quantum field theory to 
\bea
{\rm Tr}(-1)^F=\sum_E (n_B(E)-n_F(E))= n_B(0)-n_F(0)
\eea
where $n_B(E)$ and $n_F(E)$ are the number of bosonic and fermionic states with energy $E$. 
As states with vanishing energy are annihilated by $Q$,  they are not necessarily paired, i.e. $n_B(0)\neq n_F(0)$. Therefore, only the supersymmetry-preserving vacua ($E=0$) can contribute to a nonzero Witten index. The sufficient condition for unbroken SUSY is ${\rm Tr}(-1)^F\neq 0$.

\subsubsection{F-term SUSY breaking}

The O'Raifeartaigh model has three chiral superfields, $\Phi_0$, $\Phi_1$ and $\Phi_2$, with the following superpotential,
\bea
W= \mu^2\,\Phi_0 + m\, \Phi_1 \Phi_2 + g\, \Phi_0 \Phi_1 \Phi_1. 
\eea
The R-charges are $+2, 0, +2$, for $\Phi_0$, $\Phi_1$ and $\Phi_2$, respectively. 
In this model, the F-term potential is given by 
\bea
V_F = |F_0|^2 +|F_1|^2 + |F_2|^2
\eea 
where the F-terms are given by
\bea
F_0 &=& \frac{\partial W}{\partial \Phi_0} = \mu^2 + g\, \Phi_1 \Phi_1,  \nonumber \\
F_1 &=& \frac{\partial W}{\partial \Phi_1} =m \Phi_2 + 2g\, \Phi_0 \Phi_1, \nonumber \\
F_2 &=& \frac{\partial W}{\partial \Phi_2} =  m\, \Phi_1.
\eea
Then, for $\Phi_1=0$, we get $F_2=0$; $F_1=0$ for $\Phi_2=0$ and $F_0=\lambda\neq 0$. Then, SUSY is broken spontaneously.  In this case, the F-term potential is given by $V_F=|\lambda|^2$, and  $\Phi_0$ is a pseudo-flat direction, along which the R-symmetry is broken spontaneously. Actually, the SSB of R-symmetry is the sufficient condition for SUSY breaking. 

In the presence of a nonzero $F$-term, the mass  terms for scalar fields become
\bea
{\cal L}_{s-{\rm mass}}= -m^2 ( |\Phi_1|^2+ |\Phi_2|^2 ) - (g \mu^2 \Phi^2_1 +{\rm h.c.}). 
\eea
Then, the squared mass eigenvalues are $m^2_\pm=m^2\pm g\mu^2$. 
On the other hand, the fermion fields have a Dirac mass $m$,
\bea
{\cal L}_{f{\rm-mass}} = - m\psi_1\psi_2 +{\rm h.c.}.
\eea
Therefore, the scalar masses are different from the fermion mass and the mass splitting depends on the F-term.  We note that there is a sum rule for masses,
\bea
0=\sum_{s_i} (-1)^{2s_i} (2s_i+1) m^2_i = m^2_+ + m^2_- - 2 m^2
\eea
For a nonzero F-term, we obtain the SUSY transformation for $\psi_0$ as
\bea
\delta_\xi \psi_0 =  \sqrt{2} \xi F_0  =  \sqrt{2} \xi\mu^2\neq 0.
\eea
Thus, $\psi_0$ becomes a massless Goldstone fermion (Goldstino) for spontaneously broken global SUSY. The presence of a massless Goldstino is dangerous for phenomenology, but it is eaten by the spin-$\frac{3}{2}$  gravitino as longitudinal states in local supersymmetry (supergravity) \cite{supergravity}. 

{\bf Note}: There is an analogous discussion for the transformations of Goldstone bosons for global symmetry.   Suppose that $\Phi\rightarrow e^{i\alpha}\Phi$ under a global $U(1)$. Then, for $|\alpha|\ll 1$, the infinitesimal transformation is $\delta_\alpha\Phi=i\alpha \Phi$. After the $U(1)$ is spontaneously broken, we expand $\Phi=\frac{1}{\sqrt{2}}\, (v_\Phi+h+ia)\neq 0$, resulting in the $U(1)$ transformation,
\bea
\delta_\alpha a = \alpha\,  v_\Phi\neq 0.
\eea
Thus, the Goldstone $a$ transforms non-linearly under the global $U(1)$.

\subsubsection{D-term SUSY breaking}

The Fayet-Iliopoulos(FI) model is composed of two chiral superfields, $\Phi_1$ and $\Phi_2$, with opposite charges, and one $U(1)$ vector superfield $V$, with the FI D-term, with the following Lagrangian,
\bea
{\cal L}_{FI} &=& \frac{1}{4} \Big(  W^\alpha W_\alpha|_{\theta\theta}+{\rm h.c.}\Big)+ (\Phi^\dagger_+ e^{eV} \Phi_++\Phi^\dagger_- e^{-eV} \Phi_-)|_{\theta\theta{\bar\theta}{\bar\theta}}  \nonumber \\
&&+ (m\Phi_1\Phi_2|_{\theta\theta} +{\rm h.c.}) + 2\kappa^2 \, V |_{\theta\theta{\bar\theta}{\bar\theta}}. 
\eea 
In this model, the scalar potential has both F-terms and D-term as
\bea
V=|F_1|^2 +|F_2|^2 + \frac{1}{2} D^2
\eea
with
\bea
D&=& -\kappa^2 - \frac{e}{2} (A^*_1 A_1 - A^*_2 A_2), \\
F_1 &=& - mA^*_2, \\
F_2 &=& - mA^*_1. 
\eea
As a result, there is no solution to $F_1=F_2=D=0$, i.e. $V=0$. Thus, SUSY is broken spontaneously.  For instance, for $m^2>\frac{1}{2}e\kappa^2$, $A_1=A_2=0$ is the minimum, so we get $F_1=F_2=0$, but $D=-\kappa\neq 0$. 
  
In the presence of a nonzero D-term but no gauge symmetry breaking,  the scalar masses are given by
\bea
{\cal L}_{\rm s-mass} = -\Big(m^2+\frac{1}{2}e\kappa^2\Big) |\Phi_1|^2- \Big(m^2-\frac{1}{2}e\kappa^2 \Big) |\Phi_2|^2. 
\eea  
Thus, the scalar masses are shifted to $m^2_{1,2}=m^2\pm\frac{1}{2}e\kappa^2$, while the fermions have a Dirac mass $m$. Again, the scalar masses are different from the fermion mass and the mass splitting depends on the D-term. 
Then, there is a similar sum rule for masses,
\bea
0=\sum_{s_i} (-1)^{2s_i} (2s_i+1) m^2_i = m^2_1 + m^2_2 - 2 m^2
\eea
For a nonzero F-term, we obtain the SUSY transformation for $\psi_0$ as
\bea 
\delta_\xi \lambda= i \xi D =   i\xi\kappa^2 \neq 0.
\eea
Thus, the gaugino $\lambda$ becomes a massless Goldstino for spontaneously broken global SUSY. In general, there is a mixture of F-term and D-term SUSY breaking. In this case, the resulting Goldstino is also the mixture of a gaugino and a chiral fermion in the chiral superfield.

{\bf Problem}:   Find the minimum for $m^2<\frac{1}{2}e\kappa^2$ and discuss the SUSY breaking and the mass spectrum in this case.

\subsubsection{Messenger interactions}

The presence of a lighter scalar superpartner ($m^2_-$  in F-term or $m^2_2$ in D-term ) after SUSY breaking shows that SUSY must be broken in the hidden sector.  Because we have not found a charged scalar particle lighter than electron, for instance. 

In general, we can parametrize the SUSY breaking in the hidden sector by a chiral superfield $X=F_X\theta^2$ and a vector superfield $V_X=\frac{1}{2} D_X \theta^2 {\bar\theta}^2$ (or $W_{X\alpha} = \theta_\alpha D_X$). 
Then, depending on the messenger interactions, the F-term SUSY breaking leads to the scalar superpartner mass  for $\Phi$ and the gaugino mass for $V$ in the visible sector by
\bea
\int d^2\theta d^2{\bar\theta} \frac{1}{M^2_*}\, X^\dagger X \Phi^\dagger \Phi &=& \frac{|F|^2}{M^2_*}\,\phi^\dagger\phi \longrightarrow m^2_\phi= \frac{|F_X|^2}{M^2_*}=\frac{M^4_{\rm SUSY}}{M^2_*}, \\
\int d^2\theta \frac{1}{M_*}\, X W^\alpha W_\alpha &=& -\frac{F_X}{M_*}\, \lambda\lambda \, \longrightarrow m_\lambda = \frac{F_X}{M_*} =\frac{M_{\rm SUSY}}{M_*}
\eea
where $M_*$ is the mediation scale and $M_{\rm SUSY}$ is the SUSY breaking scale in the hidden sector.  

Mediation mechanisms: \\
\indent 1) Gravity mediation \cite{gravitymed}:  $M_*=M_{Pl}$. SUSY breaking is mediated by gravitational interactions, so gravitino mass and soft SUSY breaking masses are of the same order,
\bea
m_{3/2}\sim \frac{M^2_{\rm SUSY}}{M_{Pl}}\sim m_{\rm soft}.
\eea
Thus, for $M_{\rm SUSY}\sim 10^{11}\,{\rm GeV}$, we obtain $m_{2/3}\sim m_{\rm soft}\sim 1\,{\rm TeV}$. But, in this case, soft masses generically violate CP and induce FCNC, thus lack of predictive power. 

\indent 2) Gauge mediation \cite{gaugemed}: $M_*\ll M_{Pl}$.  SUSY breaking is mediated by SM gauge interactions, so there appear naturally flavor-universal and degenerate soft masses. In this case, the messeger quarks and leptons obtain masses due to direct couplings to the SUSY breaking sector.  For instance, the superpotential contains $W=\lambda_X X {\bar \Phi} \Phi$ with $\Phi, {\bar\Phi}$ being vector-like representations under the SM, and $X=M_*+\theta^2 F_X$. In this case, since soft masses are given by
\bea
m_{\rm soft} \sim \frac{\alpha}{4\pi}\, \frac{|F_X|}{M_*},
\eea 
the messenger scale can be lowered to $M_*\sim\sqrt{F_X}=10^{5-6}\,{\rm GeV}$ for $m_{\rm soft}\sim 1\,{\rm TeV}$.  
Moreover, gravitino mass is $m_{3/2}\sim 10^{-8}-10^{-6}\,{\rm GeV}$, which is a candidate for light dark matter.  

\indent 3) There are anomaly mediation \cite{anomalymed}, $Z'$ mediation,  mirage mediation, etc.  In general, SUSY breaking masses in the visible sector are a mixture of various messenger interactions.

\subsection{SUSY phenomenology}

We discuss the basics of the Minimal Supersymmetric Standard Model (MSSM)  \cite{primer,mssm,luty,textbook} and some pros and cons of the MSSM. More discussion on MSSM phenomenology such as dark matter, collider and flavor constraints, $(g-2)_\mu$, etc,  can be found elsewhere in the literature.

\subsubsection{MSSM} 

In the Minimal Supersymmetric Standard Model (MSSM), we introduce chiral superfields for SM fermions and the Higgs doublet as
\bea
{\hat Q}_i &=& {\tilde q}_i + \sqrt{2} \theta q_i + \theta^2 F_{Q_i}, \\
{\hat U}^c_i &=& {\tilde u}^c_i + \sqrt{2} \theta u^c_i + \theta^2 F_{U^c_i}, \\
{\hat D}^c_i &=& {\tilde d}^c_i + \sqrt{2} \theta d^c_i + \theta^2 F_{D^c_i}, \\
{\hat L}_i&=& {\tilde l}_i + \sqrt{2} \theta l_i + \theta^2 F_{L_i}, \\
{\hat E}^c_i &=&  {\tilde e}^c_i + \sqrt{2} \theta e^c_i + \theta^2 F_{E^c_i}, \\
{\hat H}_d &=& H_d +  \sqrt{2} \theta  {\tilde H}_d + \theta^2 F_{H_d}. 
\eea
Here, $u^c=(u^c)_L=(u_R)^c$, etc, and quark superpartners $ {\tilde q}_i ,  {\tilde u}^c_i,  {\tilde d}^c_i $ are squarks, lepton superpartners ${\tilde l}_i, {\tilde e}^c_i $  are sleptons, Higgs superpartner ${\tilde H}_d$ is Higgsino.  In order to cancel the $SU(2)_L\times U(1)_Y$ anomalies, we need to introduce one more Higgsino ${\tilde H}_u$, which makes an additional Higgs chiral multiplet,
\bea
{\hat H}_u = H_u +  \sqrt{2} \theta  {\tilde H}_u + \theta^2 F_{H_u}. 
\eea
Therefore, there are two Higgs doublets in the MSSM.
Then, the gauge-invariant superpotential is
\bea
W= y_{u,ij} {\hat Q}_i {\hat H}_u {\hat U}^c_j +  y_{d,ij} {\hat Q}_i {\hat H}_d {\hat D}^c_j +  y_{e,ij} {\hat L}_i {\hat H}_d {\hat E}^c_j  + \mu {\hat H}_u {\hat H}_d.
\eea
As a result,  the F-term potential is given by
\bea
V_F &=& \Big|\frac{\partial W}{\partial Q_i} \Big|^2+ \Big|\frac{\partial W}{\partial U^c_j} \Big|^2+\Big|\frac{\partial W}{\partial D^c_j} \Big|^2 \nonumber \\
&&+ \Big|\frac{\partial W}{\partial L_i} \Big|^2+\Big|\frac{\partial W}{\partial E^c_j} \Big|^2+ \Big|\frac{\partial W}{\partial H_u} \Big|^2+\Big|\frac{\partial W}{\partial H_d} \Big|^2 \nonumber \\
&=&y^2_{u,ij} |H_u|^2(|{\tilde u}^c_j|^2 +|{\tilde q}_i|^2) +y^2_{d,ij} |H_d|^2(|{\tilde d}^c_j|^2 +|{\tilde q}_i|^2) + y^2_{e,ij} |H_d|^2(|{\tilde e}^c_j|^2 +|{\tilde l}_i|^2)\nonumber \\
&&+ |y_{u,ij}{\tilde q}_i {\tilde u}^c_j+\mu H_d|^2+   |y_{d,ij}{\tilde q}_i {\tilde d}^c_j+y_{e,ij}{\tilde l}_i {\tilde e}^c_j+\mu H_u|^2. 
\eea
The Yukawa couplings contain those in the SM as well as new interactions in the following,
\bea
-{\cal L}_{\rm Yukawa} &=& y_{u,ij} \Big(q_i H_u u^c_j + q_i {\tilde H}_u {\tilde u}^c_j+{\tilde q}_i {\tilde H}_u u^c_j \Big) \nonumber \\
&&+  y_{d,ij} \Big( q_i H_d d^c_j  +q_i {\tilde H}_d {\tilde d}^c_j+ {\tilde q}_i {\tilde H}_d d^c_j \Big) \nonumber \\
&&+  y_{e,ij} \Big(l_i H_d e^c_j +l_i {\tilde H}_d {\tilde e}^c_j+{\tilde l}_i {\tilde H}_d e^c_j \Big)  \nonumber \\
&&+ \mu {\tilde H}_u {\tilde H}_d+{\rm h.c.}.
\eea
As a result, the Higgsinos have a Dirac mass $\mu$ and the Higgs doublets have the same masses as the Higgsino mass by SUSY.
If there is a chiral symmetry under which Higgsinos are charged, the Higgsino mass is naturally small by chiral symmetry, so the small masses of Higgs doublets are ensured by SUSY.
  
 There are spin-$\frac{1}{2}$ fermionic superpartners for the SM gauge bosons, ${\tilde g}^a$,  ${\tilde W}^i$ and ${\tilde B}$, called gluinos, winos, and bino, respectively. 
 The Lightest Supersymmetric Particle (LSP) among the neutral components of ${\tilde W}^i$, ${\tilde H}_u$, ${\tilde H}_d$, and ${\tilde B}$ (neutralinos), is a good candidate for WIMP dark matter.  The stability of LSP is ensured by $R$-parity, as will be discussed later for global symmetries in MSSM. Detailed discussion on MSSM phenomenology can be found in Ref.~\cite{textbook}.

\subsubsection{SUSY and hierarchy problem}

The top Yukawa coupling is the strongest in the SM, contributing most to the Higgs mass parameter at loop level.  In the decoupling limit with $H^0_u=\frac{1}{\sqrt{2}} h \sin\beta$ and $H^0_d=\frac{1}{\sqrt{2}} h \cos\beta$ where $\tan\beta=\frac{\langle H^0_u\rangle}{\langle H^0_d\rangle}$, the relevant couplings for the Higgs mass corrections are
\bea
{\cal L}_{\rm MSSM}\supset-\Big( \frac{1}{\sqrt{2}}\ y_t \, h {\bar t}_L  t_R+{\rm h.c.}\Big) - \frac{1}{2} y^2_t h^2(|{\tilde t}_R|^2 +|{\tilde t}_L|^2)
\eea
where ${\tilde t}_R= ({\tilde t}^c)^*$ and $y_t=y_{u,33}\sin\beta$. 
Then, the top loop corrections to the Higgs mass parameter are
\bea
(\Delta m^2_{H})_t=-\frac{N_c y^2_t}{8\pi^2}\, \Lambda^2+ \frac{3N_c y^2_t}{8\pi^2}\, m^2_t \ln \Big(\frac{\Lambda}{m_t} \Big).
\eea
whereas the stop loop contributions to the Higgs mass parameter are
\bea
(\Delta m^2_{H})_{\tilde t}=\frac{N_c y^2_t}{8\pi^2}\, \Lambda^2- \frac{N_c y^2_t}{8\pi^2}\, m^2_{\tilde t} \ln \Big(\frac{\Lambda}{m_{\tilde t}} \Big).
\eea
As a result, adding both top and stop contributions, the quadratic divergences are cancelled out, so the modified Higgs mass parameter become, for $m_{\tilde t}\gg m_t$,
\bea
\Delta m^2_{H}=- \frac{N_c y^2_t}{8\pi^2}\, m^2_{\tilde t} \ln \Big(\frac{\Lambda}{m_{\tilde t}} \Big).
\eea

SUSY is broken in nature, so we need to make superparticles heavier than the SM counterparts while maintaining the cancellation of quadratic divergences.
To this, we introduce soft SUSY breaking terms,
\bea
{\cal L}_{\rm soft}&=& -\Big(\frac{1}{2} \sum^3_{i=1} M_i \lambda_i \lambda_i +{\rm h.c.}\Big) -m^2_{H_d}|H_d|^2- m^2_{H_u}|H_u|^2 \nonumber \\
&&- m^2_{{\tilde q}, ij} |{\tilde q}_i|^2- m^2_{{\tilde u}^c, ij} |{\tilde u}^c_i|^2- m^2_{{\tilde d}^c, ij} |{\tilde u}^c_i|^2- m^2_{{\tilde l}, ij} |{\tilde l}_i|^2- m^2_{{\tilde e}^c, ij} |{\tilde e}^c_i|^2 \nonumber \\
&&-  T_{u,ij} {\tilde q}_i H_u {\tilde u}^c_j - T_{d,ij} {\tilde q}_i H_d {\tilde d}^c_j -T_{e,ij} {\tilde l}_i H_d {\tilde e}^c_j  +{\rm h.c.}.
\eea
For no FCNC, we usually choose soft masses to be flavor diagonal  by $m^2_{{\tilde q},ij}=m^2_{\tilde q} \, \delta_{ij}$, etc, and aligned by $T_{u,ij}=y_{u,ij} A_t$, etc.

\subsubsection{SUSY and vacuum stability}

The quartic terms for neutral Higgs fields come from the D-term potential as follows,
\bea
V= \frac{1}{8} (g^2+ g^{\prime 2}) (|H^0_u|^2- |H^0_d|^2)^2  
\eea 
Then, in the decoupling limit, we obtain the quartic terms as
\bea
V=\frac{1}{32} (g^2+ g^{\prime 2}) \cos^2(2\beta) h^4=\frac{1}{4} \lambda_H h^4. 
\eea
Thus, the Higgs quartic coupling is given by the electroweak gauge couplings as
\bea
\lambda_H = \frac{1}{8} (g^2+ g^{\prime 2}) \cos^2(2\beta).
\eea
Therefore, as far as superparticle masses are below the vacuum instability scale, the Higgs quartic coupling is maintained to be positive all the way to the unification scale. 

For the quartic coupling at tree level, the Higgs boson mass  is given by
\bea
m_h=\sqrt{2\lambda_H} \, v = \frac{1}{2} \sqrt{g^2+ g^{\prime 2}} \cos(2\beta) \, v \leq m_Z.
\eea
After including the top loop corrections to the Higgs quartic coupling, we can accommodate the correct Higgs boson mass with a shift in the quartic coupling,
\bea
\Delta\lambda_H = \frac{3m^4_t}{4\pi^2 v^4}\, \bigg[\ln \Big(\frac{m^2_{\tilde t}}{m^2_t}\Big) + \frac{X^2_t}{m^2_{\tilde t}}\Big( 1-\frac{1}{12}  \frac{X^2_t}{m^2_{\tilde t}}\Big) \bigg]
\eea
with $X_t=A_t -\mu \cot\beta$. In this case, the required stop masses are at least multi-TeV scales.

{\bf Problem}: Obtain the stop contributions to the Coleman-Weinberg potential for the Higgs boson and identify the corrections to the Higgs quartic coupling.

\subsubsection{Global symmetries in MSSM}

We discuss the $R$-symmetry as  the unique global symmetry in supersymmetric models and the fate of $B$ and $L$ numbers in MSSM.

\underline{$R$-symmetry}

The SUSY algebra is extended by the R-symmetry generator, which does not commute with the SUSY operators,
\bea
[Q_\alpha,R]=-Q_\alpha, \quad\quad [{\bar Q}_{\dot\alpha},R]= + {\bar Q}_{\dot\alpha}.
\eea
Then, from 
\bea
({\bar Q}_{\dot\alpha} R - R{\bar Q}_{\dot\alpha}) |B\rangle=\sqrt{E} (r_B-r_F) |F\rangle = {\bar Q}_{\dot\alpha}  |B\rangle=\sqrt{E}  |F\rangle,
\eea
where $r_B, r_F$ are the $R$-charges of boson and fermion in a chiral multiplet, 
we find that $r_B=r_F+1$. The $R$-symmetry is a global symmetry of supersymmetric models and can be also a local symmetry in supergravity.  
The $R$-symmetry or $U(1)_R$ symmetry can be broken to discrete subgroups by chiral anomalies or compactification of extra dimensions on orbifolds having discrete internal symmetries.
The discrete remnants of the continuous $R$-symmetry such as $Z_{2R}$ ($R$-parity) or $Z_{4R}$ are important for the MSSM phenomenology.

Since the superfield $\Phi=A+\sqrt{2} \theta \psi +\theta^2 F$ has the same $R$-charge as the boson, the Grassmannian variable $\theta$ has $R$-charge $+1$, so the auxiliary field $F$ has $R$-charge $r_F=r_B-2$. The $R$-symmetry transformation for a chiral superfield $\Phi$ with $R$-charge $r$ is
\bea
R \,\Phi(\theta,x)&=& e^{ir\alpha} \Phi(e^{-i\alpha}\theta,x), \\
R\, \Phi^\dagger({\bar\theta},x)&=& e^{ir\alpha} \Phi^\dagger(e^{i\alpha}{\bar\theta},x),
\eea
leading to the $R$-symmetry transformations for component fields,
\bea
A&\longrightarrow& e^{ir\alpha} A, \\
\psi & \longrightarrow & e^{i(r-1)\alpha} \psi, \\
F &\longrightarrow & e^{i(r-2)\alpha} F.
\eea
The R charge of a product of superfields is the sum of the individual R charges.
On the other hand, vector superfields have 0 $R$-charge, because they are real, so $R\, V(\theta, {\bar\theta}, x)=V(e^{-i\alpha}\theta, e^{i\alpha} {\bar\theta},x)$. 
So, in Wess-Zumino gauge, the $R$-transformations for components are
\bea
v_m\rightarrow v_m,\quad \lambda\rightarrow e^{i\alpha} \lambda, \quad D\rightarrow D.
\eea

The $R$-parity is a $Z_{2R}$ discrete symmetry with $\alpha=\pi$.
In MSSM, we take $r=1$  for quark and lepton chiral multiplets and $r=0$ for Higgs chiral multiplets. Then, the $R$-parities for chiral superfields in MSSM are assigned as
\bea
{\hat Q}_i, \,\,{\hat U}^c_i, \,\,{\hat D}^c_i,\,\, {\hat L}_i,\,\, {\hat E}^c_i &:&  \quad Z_{2R}=-1, \\
{\hat H}_{u,d} &:& \quad Z_{2R}=+1.
\eea
so the $R$-charges for component fields are
\bea
q_i, \,\, u^c_i,  \,\, d^c_i, \,\, l_i, \,\, e^c_i, \,\, H_{u,d}, \,\, v_{m,i} &:& \quad Z_{2R}=+1, \\
{\tilde q}_i, \,\,  {\tilde u}^c_i, \,\,  {\tilde d}^c_i, \,\, {\tilde l}_i, \,\, {\tilde e}^c_i, \,\, {\tilde H}_{u,d}, \,\, \lambda_i &:&  \quad Z_{2R}=-1.
\eea
The $R$-parity is related to the matter parity $P_M=(-1)^{3(B-L)}$ by $Z_{2R}=(-1)^{2S}P_M$  with $S$ being the spin.

\underline{Lepton and baryon numbers}

There are additional gauge-invariant terms in the effective superpotential in MSSM up to dimension-5 operators, given by
\bea
\Delta W= W_{\slashed{R}}+ \frac{1}{M_{Pl}}\Big( c {\hat L} {\hat H}_u {\hat L} {\hat H}_u  + \lambda_1 {\hat Q}{\hat Q}{\hat Q}{\hat L}+ \lambda_2 {\hat U}^c {\hat U}^c {\hat D}^c {\hat E}^c  +\cdots \Big).
\eea
with
\bea
 W_{\slashed{R}}=\mu' {\hat H} {\hat L}_i +\lambda_{ijk} {\hat L}_i {\hat L}_j {\hat E}^c_j + \lambda'_{ijk} {\hat L}_i {\hat Q}_j {\hat D}^c_k + \lambda^{\prime\prime}_{ijk} {\hat U}^c_i {\hat D}^c_j {\hat D}^c_k. 
\eea
Here, $ W_{\slashed{R}}$ breaks the $R$-parity, violating lepton or baryon numbers at dimension-4 level, unlike in the SM.   The dimension-5 operators preserve the $R$-parity, but they break baryon and/or lepton numbers.  $ {\hat L} {\hat H}_u {\hat L} {\hat H}_u$ is the supersymmetric Weinberg operator violating the lepton number. $\lambda_{1,2}$ couplings are induced by colored Higgsinos in supersymmetric $SU(5)$ GUTs, but they are most constrained by proton stability for low-energy supersymmetry.

For TeV-scale SUSY, the phenomenological constraints on the additional couplings are the following:
neutrino masses lead to
\bea
\mu'\lesssim 10^{-21}\, M_{Pl},
\eea
proton stability requires
\bea
|\lambda' \lambda^{\prime\prime}| \lesssim 10^{-26}, \quad |\lambda'\lambda_3|\lesssim 10^{-10}, \quad |\lambda_1|\lesssim 10^{-7}, \quad |\lambda_2|\lesssim 10^{-7}.
\eea

\subsubsection{Gauge coupling unification}

The gauge couplings in the MSSM run in energy by the RG equations,
\bea
\frac{d\alpha^{-1}_i}{d\ln\mu} = - \frac{b_i}{2\pi}, \qquad i=1,2,3,
\eea
where the corresponding beta function coefficients are given by  $b_i= (\frac{33}{5},1,-3)$ with $i=1,2,3$.   Then, the gauge couplings are unified at $M_{\rm GUT}\sim 10^{16}\,{\rm TeV}$ for TeV-scale supersymmetry.

{\bf Problem}: Evaluate the beta function coefficients for gauge couplings in MSSM.

\subsubsection{Problems in MSSM}

We address phenomenological problems in MSSM and some of solutions beyond gauge symmetries.

\underline{Proton decay}

The $R$-parity forbids the dangerous renormalizable $B/L$ violating terms in the superpotential, $W_{\slashed{R}}$. 
However, the Weinberg operator for neutrino masses as well as the dimension-5 baryon number violating couplings, $\lambda_1$ and $\lambda_2$, are allowed.  The latter couplings would induce the proton decay, $p\rightarrow K^+{\bar \nu}$, by superpartner loops, so we need a symmetry beyond the $R$-parity for proton stability.   From the proton lifetime, 
\bea
\tau(p\rightarrow K^+ \nu)\sim \frac{(16\pi^2)^2M^2_P m^2_{\rm soft}}{\lambda^2_1 m^5_p}>5.9\times 10^{33}\,{\rm yrs},  
\eea
we obtain the bound, $(16\pi^2 M_{Pl} m_{\rm soft}/\lambda_1)^{1/2}\gtrsim 10^{15}\,{\rm GeV}$.  Thus, for $m_{\rm soft}\sim 1\,{\rm TeV}$, we need a very small coupling for the dimension-5 operator, $\lambda_1\lesssim 10^{-7}$.

\underline{The $\mu$ problem}

The $\mu$ term is gauge invariant and $R$-parity invariant. 
In the presence of the $\mu$ term, Higgs bosons get the same supersymmetric mass as Higgsinos as well as SUSY breaking masses, as follows,
\bea
m^2_1= |\mu|^2 +m^2_{H_d}, \quad
m^2_2= |\mu|^2 +m^2_{H_u}.
\eea
Then, the condition for electroweak symmetry breaking requires the $\mu$ term to be related to the $Z$-boson mass by
\bea
|\mu|^2 = -\frac{1}{2} m^2_Z + \frac{m^2_{H_d}-m^2_{H_u}\tan^2\beta}{\tan^2\beta-1}.
\eea
Thus, in order not to introduce a fine-tuning, we need $\mu$ to be about the weak scale. 
This is the $\mu$ problem. 

A simple solution is to introduce a chiral $U(1)$ symmetry under which $H_u H_d$ is charged such that the $\mu$ term vanishes at tree-level and it is induced after the $U(1)$ symmetry is broken spontaneously. The examples are $U(1)_{\rm PQ}$ and $U(1)_R$. The discrete subgroups of $U(1)_R$ larger than $Z_{2R}$ can solve the $\mu$ problem as well as the proton instability problem in MSSM.  

The $Z_{4R}$ symmetry is consistent with $SU(5)$ unification, assigning  $R$-charge $+1$ for quark and lepton superfields and $R$-charge $0$  for Higgs superfields  \cite{rsym}. In this case, we can show that the $\mu$ term and the dangerous dimension-5 proton decay operators are absent at tree level while the Weinberg operator for neutrino masses are allowed.  The $Z_{4R}$ symmetry can be broken by non-perturbative effects such as gaugino condensation or the VEVs of singlet chiral superfields (in NMSSM \cite{nmssm}), leading to a desirably small $\mu$ term.
For gauged $R$ symmetry, the corresponding massive vector multiplet can be responsible for mediation of SUSY breaking \cite{U1Rmed}.

\underline{Flavor and CP problems}

The general soft breaking mass terms in gravity-mediation, such as  $m_{{\tilde q},ij}$ and $T_{u,ij}$, etc, would lead to dangerous flavor and CP violations beyond the SM. This is called the SUSY flavor problem. Thus, new flavor and CP violating couplings due to soft masses are constrained by FCNC processes and electric dipole moments mostly for the first two generations. 

The simple solution to the SUSY flavor problem is to take $m_{{\tilde q},ij}$ to be diagonal as in gauge mediation and  $T_{u,ij}$ to be aligned as the corresponding Yukawa couplings in the SM. Another solution is to take the first two generation sfermions to be much heaver than weak scale to satisfy the flavor constraints.

\section{Alternatives to SUSY}

We discuss alternative solutions to the hierarchy problem, based on extra dimensions, clockwork mechanism, relaxion, twin Higgs and four-form flux models.

\subsection{Extra dimensions}

Extra spatial dimensions are a general consequence of the consistency of string theories defined in 10D or 11D.   Since extra dimensions are not seen, they must be curled up into small sizes, needing the process of the so called compactification.  When one extra dimension is compactified on a circle ($S^1$) with radius $R$, we can make a Fourier expansion of a 5D massless scalar field as
\bea
\Phi(x,y)= \sum_{n=\infty}^\infty e^{in y/R} \phi_n(x),
\eea
where $(\Box+m^2_n)\phi_n(x)=0$ with $m_n= \frac{n}{R}$ and $n$ being integer.
Then, the Kaluza-Klein modes $\phi_n$ have 4D mass $m_n$ due to the momentum in the extra dimension.  Such a Fourier expansion in more than 5D spacetime is straightforward. 

The Gauss law for gravity with $n$ extra dimensions shows us that the gravity force is given by
\bea
F_g=- \frac{G_{4+n} m_1 m_2}{r^{2+n}}.
\eea 
When $n$ extra dimensions have a radius $R$, at the distance scale larger than $R$, the gravity force becomes the Newtonian gravity in 4D as
\bea
F_g=- \frac{G_{4+n} m_1 m_2}{R^n r^{2}}= - \frac{G_N m_1 m_2}{r^{2}}.
\eea
with
\bea
G_N = \frac{G_{4+n}}{R^n}.  
\eea
Then, the weakness of 4D gravity  can be explained for large extra dimensions, due to the volume suppression in the extra dimensions \cite{LED}. In terms of Planck mass scales in 4D and higher dimensions by $G_N=8\pi M^2_P$ and $G_{4+n}=8\pi (M_{4+n})^{2+n}$, respectively, we obtain $M^2_P= (M_{4+n})^{2+n} R^n$.  Therefore, we can explain the hierarchy between the Planck scale and the weak scale, $M_{Pl}\gg M_{4+n}\sim m_h$, at the expense of the new hierarchy $R^{-1}\ll M_{4+n}$. In this case, extra dimensions are called flat or factorizable or unwarped, in the sense that they are factorized from 4D spacetime.

In the case of warped extra dimensions, extra dimensions are not factorized from 4D spacetime \cite{WED}. The metric in 5D spacetime with one warped extra dimension is given by
\bea
ds^2 =\omega^2(y) \eta_{\mu\nu} dx^\mu dx^\nu- dy^2. 
\eea
Here, $\omega(y) $ is the warp factor. 
When the 5D bulk cosmological constant is negative as 
\bea
\Lambda_b \equiv -6k^2 M^3_5
\eea
 with $M_5$ being the 5D Planck mass and there is a $Z_2$ symmetry under $y\rightarrow -y$, the warp factor takes the form, 
 \bea
 \omega(y)=e^{-k|y|}. 
 \eea
 In this case, in order to satisfy the Einstein equations in 5D, we need to introduce two brane tensions localized at $y=0$ and $y=\pi R$ by
\bea
\Lambda_1 = -\Lambda_2 =\sqrt{-6M^3_5 \Lambda_b}=6k M^3_5.
\eea
Due to the warped factor, for a fixed 4D proper distance, the 4D coordinate distance changes along the extra dimension: it becomes larger (or the inverse distance becomes smaller) as $y$ increases.  For this reason, $y=0$ is the UV brane and $y=\pi R$ is the IR brane. 
When the Higgs doublet is localized on the IR brane, the corresponding action is
\bea
S_H=\int d^4x \sqrt{-h}  \bigg(h^{\mu\nu} (D_\mu H)(D_\nu H)^\dagger- \lambda_H \Big(|H|^2-\frac{v^2}{2}\Big)^2  \bigg).
\eea
Here, $h_{\mu\nu}=e^{-2k\pi R}\,\eta_{\mu\nu}$ is the induced metric on the IR brane, and $h^{\mu\nu}=e^{2k\pi R}\,\eta^{\mu\nu}$ is the inverse.
Then, the above action becomes
\bea
S_H&=&\int d^4x  \bigg( e^{-2k\pi R} \eta^{\mu\nu} (D_\mu H)(D_\nu H)^\dagger-  e^{-4k\pi R} \lambda_H \Big(|H|^2-\frac{w^2}{2}\Big)^2  \bigg) \nonumber \\
&=&\int d^4x  \bigg( \eta^{\mu\nu} (D_\mu H)(D_\nu H)^\dagger-  e^{-4k\pi R} \lambda_H \Big(|H|^2-\frac{v^2}{2}\Big)^2 \bigg)
\eea
where a field definition is made by ${\tilde H}=e^{-k\pi R}H $, and the electroweak scale is
\bea
v= e^{-k\pi R}\, w.
\eea
Therefore, for $w\sim M_5\sim M_{Pl}$, we can obtain the hierarchy between the Planck scale and the weak scale for $k\pi R=37$, which needs a mild hierarchy between $k$ and $R^{-1}$. 
In this case, the 4D Planck mass is related to the 5D Planck mass by
\bea
M^2_P= \frac{M^3_5}{k} \Big(1-e^{-2k\pi R} \Big) \approx  \frac{M^3_5}{k}. 
\eea
Therefore, $M_5$ and $k$ can be of order the 4D Planck scale. In this case, the weakness of gravity can be understood because the zero mode of graviton is localized on the UV brane.

\subsection{Clockwork theory}

The clockwork mechanism is to obtain hierarchically small couplings in models with multiple symmetries due to the localization in the field space \cite{clockwork}. It can address the hierarchy problem and the flavor problem, etc, in the SM, by introducing multiple copies of the SM particles. 
The clockwork setup can be shown to be realized in a 5D dilaton background with warped extra dimension. 

We consider $N+1$ global symmetries, $U(1)_0\times U(1)_1\times\cdots \times U(1)_{N}$, in 4D. We introduce complex scalar fields $\Phi_0, \Phi_1,\cdots,\Phi_N$, that carry global $U(1)$ charges for the nearest neighbor $U(1)$'s such as $(q,1)$.
Then, the potential terms are given by
\bea
V(\Phi)=\sum_{j=0}^{N} \Big(m^2 |\Phi_j|^2+\lambda |\Phi_j|^4 \Big)+\sum_{j=0}^{N-1} \frac{\lambda'}{2} \Phi^q_j \Phi^\dagger_{j+1} +{\rm h.c.}
\eea
Here, we took the universal masses and the quartic couplings as well as the $U(1)$ breaking terms, but they can be generalized to general parameters. 

We ignore the effects of $U(1)$-breaking terms on the VEVs of complex scalar fields, and take  the radion modes to be frozen such that $\Phi_j= \frac{f}{\sqrt{2}}\, e^{i\pi_j/f}$. Then, the effective potential for Goldstones $\pi_j$ is
\bea
V(\pi) =\sum_{j=0}^{N-1} \lambda' f^{q+1} \cos\Big(\frac{q\pi_j - \pi_{j+1}}{f} \Big).  \label{effpot}
\eea
Then, the quadratic potential for $\pi_j$ becomes
\bea
V_2(\pi) = \sum_{j=0}^{N-1} \frac{1}{2} m^2  \Big(q\pi_j-\pi_{j+1}\Big)^2
\eea
with $m^2\equiv \lambda' f^{q-1}$. Then, the mass matrix for scalar fields is given by
\bea
M^2_\pi= m^2 \left(\begin{array}{cccccc} q^2 & -q & 0 &  \cdots &  & 0 \\  -q & 1+q^2 & -q  & \cdots & & 0 \\ 0 & -q  & 1+q^2 & \cdots &  & 0  \\   \vdots & \vdots & \vdots & \ddots & & \vdots  \\  & & &  & 1+q^2 & -q \\ 0 & 0 &  0 & \cdots & -q & 1 \end{array}  \right).
\eea
The effective potential (\ref{effpot}) respects the unbroken shift symmetry, under $\pi_j\rightarrow \pi_j+ c \, q^j $ with $c$ being constant, which is the unbroken $U(1)$ symmetry.
 
 As.a result, the zero mode of the scalar clockwork is  given  by
\bea
{\tilde \pi}_0(x)=\sum_{j=0}^N a_{j0} \pi_j(x)
\eea
where $a_{j0}=N_0 \, q^j$ with $N_0=\sqrt{(q^2-1)/(q^{2(N+1)}-1) }$.
On the other hand, the massive modes of the scalar clockwork are also given by
\bea
{\tilde \pi}_k(x)=\sum_{j=0}^N a_{jk} \pi_j(x),\quad k=1,2,\cdots, N,
\eea
with the mass eigenvalues being
\bea
M^2_k=m^2 \Big( 1+q^2-2q \cos\frac{k\pi}{N+1}\Big)\equiv m^2\lambda_k, \label{mass}
\eea
and the wave functions being 
\bea
a_{jk}= N_k \left[q\sin\Big(\frac{jk\pi}{N+1} \Big)-\sin\Big(\frac{(j+1)k\pi}{N+1} \Big) \right], \quad N_k=\sqrt{\frac{2}{(N+1)\lambda_k}}.  \label{wf}
\eea
Massive modes have an overall mass gap $m$ from the zero mode and have the squeezed mass spectrum, $\delta M_k/M_k\sim 1/N$, for a large $N$. 
We note that the interacting gauge fields  are invertible to get 
\be
\pi_j(x)=\sum^N_{i=0} a_{kj} {\tilde \pi}_k(x), \quad j=0,1,2,\cdots,N.  \label{flavor}
\ee
Thus, the zero mode is localized toward the site at $j=N$  for $q>1$, so it has position-dependent couplings to external fields, in particular, suppressed couplings to external fields localized at $j=0$ by 
\bea
{\cal L}_{\rm int}=\frac{1}{f}\,\pi_0\, {\cal O}_{\rm ext}=\frac{1}{f_{\rm eff}}\,{\tilde \pi}_0 \, {\cal O}_{\rm ext} +\cdots, \qquad f_{\rm eff} \equiv f q^N \gg f.
\eea

At the quadratic level for Goldstones, the scalar clockwork is equivalent to a massless scalar field in 5D dilaton background. The 5D coordinate is given by $y=j a$ with $a$ being the lattice distance. In the continuum limit  we take $a\rightarrow 0$ and $N\rightarrow \infty$ while $\pi R= N\,a $ being finite.  From the identification, $\pi_j(x)=e^{ky}\phi(x,y)$,  with $k=(q-1)/(qa)$ and $q^N=e^{k\pi R}$, the corresponding 5D Lagrangian \cite{CW2} is given by 
\bea
{\cal L}_{5D} =\int^{\pi R}_0 dy\, e^S\, \frac{1}{2} \partial_M \phi \, \partial^M\phi +\int^{\pi R}_0 dy\,\delta(y)\, e^{\frac{1}{2}S}\, \phi\, {\cal O}_{\rm ext}
\eea
where the dilaton background is given by $S=2k |y|$.  The nontrivial dilaton background is supported by the 5D warped geometry with the metric \cite{5Ddilaton}, 
\bea
ds^2= e^{\frac{4}{3}k|y|} (\eta_{\mu\nu} dx^\mu dx^\nu -dy^2) , 
\eea
and nonzero brane tensions, $\Lambda_0=-\Lambda_\pi=-4kM^3_5$  at $y=0$ and $y=\pi R$, respectively, where $M_5$ is the 5D Planck scale. 
Then, the zero mode of $\phi(x,y)$ is constant such that $\pi_j(x)\sim e^{ky}$ is localized at $y=0$ in the continuum limit.
In this case, the 4D Planck mass is related to the 5D Planck mass by
\bea
M^2_P= \frac{M^3_5}{k} \Big(e^{2k\pi R}-1 \Big) \approx  \frac{1}{3} M^3_5\, L_5 \, e^{\frac{4}{3}k\pi R}
\eea
with $L_5=\int dy\, \sqrt{-g_{55}}=\frac{3}{k}\,(e^{\frac{2}{3}k\pi R}-1)$ being the proper radius of the extra dimension. In this case, for $k\sim M_5$, we obtain the hierarchy of scales,
\bea
\frac{M^2_5}{M^2_P}  \ll  \frac{(L_5)^{-1}}{M_5}\ll 1. 
\eea
Therefore, when the Higgs mass parameter is of order $M_5\sim k$, we can solve the hierarchy problem by the warp factor.

\subsection{Relaxion mechanism}

The relaxion mechanism is to address the hierarchy problem from the cosmological evolution of the Higgs mass parameter, instead of relying on the symmetries to protect the Higgs mass \cite{relaxion}. This is in a similar spirit as the axion solution to the strong CP problem \cite{axion1,axion2}. 
 
We consider a relaxion scalar (or axion) $\phi$ with the coupling to the SM Higgs, in the following,
\bea
{\cal L} = -(-M^2+g\phi) |H|^2 - V_\phi  +\frac{1}{32\pi^2} \frac{\phi}{f} G_{\mu\nu} {\tilde G}^{\mu\nu}
\eea
with 
\bea
V_\phi = g M^2 \phi + g^2 \phi^2+\cdots.
\eea
Here, $g$ is a dimensionful parameter, $M$ is the cutoff scale, and the relaxion potential is valid for $\phi\lesssim M^2/g$. 
Then, after QCD condensation, we obtain the effective potential as
\bea
V_{\rm eff} = (-M^2 + g\phi) |H|^2 + (gM^2 \phi+ g^2 \phi^2+\cdots)  +\Lambda^4 \cos\Big(\frac{\phi}{f}\Big)
\eea
with $\Lambda^4\sim f^2_\pi m^2_\pi\sim(0.1\,{\rm GeV})^4$. 

There are several conditions to be fulfilled for the relaxion mechanism to work: \\
\indent 1) Slow-roll of relaxion: $M_{Pl}\, V'_\phi/V_I<1$ where $V_I$ is the inflaton potential  and $H_I=V_I/3M^2_P$. \\
\indent From $V_I=3M^2_P H^2$ and $V'_\phi\sim gM^2$, this leads to 
\bea
g< \frac{H^2_I M_{Pl}}{M^2}.
\eea
\indent 2) Sufficient inflation for $\Delta \phi\gtrsim M^2/g$: \\
\bea
\Delta\phi= {\dot\phi} \Delta t =  {\dot\phi}\, \frac{N}{H_I} \sim \frac{V'_\phi}{H^2_I}\, N\sim \frac{gM^2}{H^2_I}\, N\gtrsim \frac{M^2}{g},
\eea
\indent\indent results in the bound on the number of efoldings,
\bea
N\gtrsim \frac{H^2_I}{g^2}.
\eea
\indent 3) Conditions on the Hubble scale $H_I$: \\
\indent \indent i) $V_I> V_\phi$ gives rise to the lower bound, $M^2_PH^2_I > M^4$ or 
\bea
H_I> \frac{M^2}{M_{Pl}}. \label{H1}
\eea
\indent \indent ii) Classical rolling of $\phi$ during one Hubble time leads to the upper bound:
\bea
\frac{\delta \phi}{\Delta\phi/N}<1,\quad {\rm i.e.} \quad  \frac{H_I}{V'_\phi/H^2_I} <1
\eea 
\indent\indent leading
\bea
H_I< (gM^2)^{\frac{1}{3}}. \label{H2}
\eea
\indent \indent iii) Barriers from QCD phase transition form in Hubble volume: \\
\bea
H_I<\Lambda. \label{H3}
\eea
\indent \indent Thus, considering eqs.~(\ref{H1})-(\ref{H3}) together, we get the bounds on $H_I$,
\bea
\frac{M^2}{M_{Pl}}<H_I< {\rm min} \Big\{(gM^2)^{\frac{1}{3}}, \Lambda\Big\}.
\eea
The initial condition for relaxion is $m^2_H=-M^2+g\phi>0$ in the unbroken phase and $\phi$ slow-rolls  during inflation. When $m^2_H<0$ developing $\langle H\rangle\neq 0$, the QCD potential becomes nonzero because $\Lambda^4\propto m_u \Lambda^3_{\rm QCD}\sim \langle H\rangle$.  Then, the large barries stop the rolling of $\phi$ shortly after $\phi=M^2/g$  Slow-rolling of $\phi$ stops when $V'_{\rm QCD}+V'_{\phi}=0$, that is,
\bea
\frac{\Lambda^4}{f} \sim gM^2. \label{stop}
\eea
Therefore, as $(gM^2)^{1/3}\sim (\Lambda^4/f)^{1/3}=\Lambda (\Lambda/f)^{1/3}\ll \Lambda$ for $f\gg \Lambda$, the bound on $H_I$ becomes
\bea
\frac{M^2}{M_{Pl}}< \Big( \frac{\Lambda^4}{f}\Big)^{\frac{1}{3}}.  \label{cutoff}
\eea
As a result, the cutoff scale is bounded as
\bea
M< \bigg(\frac{\Lambda^4M^3_P}{f} \bigg)^{\frac{1}{6}}\sim 10^7\,{\rm GeV} \bigg(\frac{10^9\,{\rm GeV}}{f} \bigg)^{\frac{1}{6}}.
\eea 
From eq.~(\ref{stop}) with $M=10^7\,{\rm GeV}$ and $f= 10^9\,{\rm GeV}$, we get  $g\sim 10^{-27}\,{\rm GeV}$, so $H_I< 10^{-5}\,{\rm GeV}$ and $N\gtrsim 10^{44}$.

We note that when the relaxion is stabilized by the QCD potential, the minimum value for $\phi$ is determined from $\sin \Big(\frac{\phi}{f}\Big)\sim gM^2 f /\Lambda^4$, so $\langle \phi/f\rangle={\cal O}(1)$ for $gM^2\sim \Lambda^4/f$. But, in order for the relaxion to be a solution to the strong CP problem, we would need $gM^2< 10^{-10} \Lambda^4/f$. In this case, instead of eq.~(\ref{cutoff}), the cutoff scale is bounded to
\bea
M<  \bigg(10^{-10}\,\frac{\Lambda^4M^3_P}{f} \bigg)^{\frac{1}{6}}\sim 100\,{\rm TeV}  \bigg(\frac{10^9\,{\rm GeV}}{f} \bigg)^{\frac{1}{6}}. 
\eea
In this case, for $M=100\,{\rm TeV}$ and $f=10^9\,{\rm GeV}$, we get $g=10^{-33}\,{\rm GeV}$, so $H_I<10^{-8}\,{\rm GeV}$ and $N\gtrsim 10^{50}$.

We also remark that the bound on the cutoff scale can be relaxed if the condition for classical rolling  of $\phi$, eq.~(\ref{H2}), is ignored. In this case, the upper bound on the cutoff scale comes from eq.~(\ref{H3}), so
\bea
M< \sqrt{\Lambda M_{Pl}} \sim 10^8\,{\rm GeV}. 
\eea
In this case, keeping the solution to the strong CP problem by $gM^2< 10^{-10} \Lambda^4/f$, we obtain $g=10^{-39}\,{\rm GeV}$ for $M=10^8\,{\rm GeV}$ and $f= 10^9\,{\rm GeV}$, so $H_I<\Lambda=0.1\,{\rm GeV}$ and $N\gtrsim 10^{76}$. 

There are extended discussions on two-field relaxion scenarios \cite{relaxion2} and supersymmetric 
UV completion of relaxion models \cite{relaxion-susy}.

\subsection{Twin Higgs models}

Twin Higgs models are to introduce a mirror copy of the SM gauge groups $[SU(2)_A\times U(1)_A]\times [SU(2)_B\times U(1)_B]$ as well as the SM matter content \cite{twinHiggs}. $SU(3)_A\times SU(3)_B$ factor can be also included. Moreover, the SM and mirror gauge symmetries can be unified into $SU(4)\times SU(6)$.  
Due to the $Z_2$ symmetry between the SM and the mirror SM, $H_A \leftrightarrow H_B$, the Higgs potential takes 
\bea
V= m^2(|H_A|^2+|H_B|^2) +\lambda (|H_A|^4+|H_B|^4) +2\lambda' |H_A|^2 |H_B|^2.
\eea
For $\lambda'=\lambda$, there is a full $SU(4)$ global symmetry in the potential, which is gauged by $SU(2)_A\times SU(2)_B$.

The  $SU(4)$ global symmetry is broken spontaneously by the mirror Higgs VEV $f$ as
\bea
\left(\begin{array}{c} H_A  \\  H_B\end{array}\right)= e^{i h^a t^a/f}\, \left(\begin{array}{c} 0 \\ 0  \\ 0 \\  f+\rho\end{array}\right)
\eea
where $t^a$ are seven broken generators belonging to $SU(4)/SU(3)$.  
Then, six of the pseudo-Goldstone bosons among $h^a$ are eaten by the massive electroweak gauge bosons in the SM and mirror partners, while there is a light SM Higgs boson in the low energy. The $SU(2)_B\times U(1)_B$ gauge bosons receive masses, $m^2_{W_B}=g^2_B f^2/2$ and $m^2_{Z_B}=(g^2_B +g^{\prime 2}_B) f^2/2$, while the mirror electromagnetism would remain unbroken. 
We can show the pseudo-Goldstones explicitly in the following representation,
\bea
\left(\begin{array}{c} H_A  \\  H_B\end{array}\right)= {\rm exp}\left[\frac{i}{f} \left(\begin{array}{cccc} 0 & 0 & 0 & h_1 \\  0 & 0 & 0 & h_2 \\ 0 & 0 & 0 & h_3 \\ h^*_1 & h^*_2 & h^*_3 & h_0 \end{array} \right) \right] \left(\begin{array}{c} 0 \\ 0  \\ 0 \\  f+\rho\end{array}\right) 
\eea
where $f=\sqrt{-m^2/2\lambda}$ is the $SU(4)$ breaking scale, and $h_1, h_2, h_3$ are complex scalar fields,  and $h_0$ is real.
After removing $h_3, h_0$ by $SU(2)_B\times U(1)_B$ gauge transformations, we are left with
\bea
\left(\begin{array}{c} H_A  \\  H_B\end{array}\right)=(f+\rho)\left(\begin{array}{c}  i\frac{h_1}{|h|} \sin\frac{|h|}{f} \\i\frac{h_2}{|h|} \sin\frac{|h|}{f}   \\ 0 \\  \cos\frac{|h|}{f}\end{array} \right)
\eea
In the decoupling limit of the mirror Higgs partner $\rho$, we have
\bea
\left(\begin{array}{c} H_A  \\  H_B\end{array}\right)= f \left(\begin{array}{c}  i\frac{h_1}{|h|} \sin\frac{|h|}{f} \\i\frac{h_2}{|h|} \sin\frac{|h|}{f}   \\ 0 \\  \cos\frac{|h|}{f}\end{array} \right)\approx
 \left(\begin{array}{c} ih_1+\cdots  \\  ih_2+\cdots \\ 0 \\  f-\frac{|h|^2}{2f}+\cdots \end{array} \right)\equiv  \left(\begin{array}{c} \phi^++\cdots  \\  \phi^0+\cdots \\ 0 \\  f-\frac{|\phi^0|+|\phi^+|^2}{2f}+\cdots \end{array} \right)
\eea
where $|h|=\sqrt{|h_1|^2+|h_2|^2}$ and $H^T=(\phi^+,\phi^0)^T$ is the remaining $SU(2)_A$ doublet, being identified as the SM Higgs doublet.

From
\bea
H^\dagger_A H_A &=& f^2 \sin\frac{|h|}{f} = h^\dagger h -\frac{(h^\dagger h)^2}{3f^2}+\cdots,  \label{HA-exp} \\
 H^\dagger_B H_B &=& f^2 \cos\frac{|h|}{f} =  f^2 - h^\dagger h + \frac{(h^\dagger h)^2}{3f^2}+\cdots, \label{HB-exp}
 \eea
 the $SU(4)$-invariant potential does not contain the SM Higgs doublet, but the $SU(4)$ symmetry is broken by gauge interactions as well as Yukawa couplings, so there appears a nonzero potential for the SM Higgs doublet. 
  
The Yukawa couplings for third generation quarks and their mirror partners are
\bea
{\cal L}_Y=-y_A H_A {\bar Q}_{AL} t_{AR} - y_B  H_B {\bar Q}_{BL} t_{BR} +{\rm h.c.} 
\eea
where $y_A=y_B$ is taken due to the $Z_2$ symmetry. Inserting the expanded form of the Higgs doublets in the above, we get the mass for mirror top quark as $m_{t_B}= f y_B$ and the Yukawa couplings for the SM doublet as
\bea
{\cal L}_Y&=&- y_A H  {\bar Q}_{AL} t_{AR}+ \frac{y_A}{2f} \, |H|^2\, {\bar t}_{BL} t_{BR} +{\rm h.c.}  \nonumber \\
&=& - y_A \phi^0\, {\bar t}_A t_A +  \frac{y_A}{2f}\, |\phi^0|^2\, {\bar t}_B t_B +\cdots.
\eea
As a consequence, the one-loop corrections from top and mirror top quarks to the SM Higgs mass parameter are
\bea
\Big(\Delta m^2_H\Big)_{t_A}&=& -\frac{N_c y^2_A}{8\pi^2}\, \Lambda^2 +\frac{3N_c y^2_A}{8\pi^2}\,m^2_{t_A}\ln \Big(\frac{\Lambda}{m_{t_A}} \Big), \nonumber \\
\Big(\Delta m^2_H\Big)_{t_B}&=& \frac{N_c y^2_A}{8\pi^2}\, \Lambda^2 -\frac{N_c y^2_A}{4\pi^2}\,m^2_{t_B}\ln \Big(\frac{\Lambda}{m_{t_B}} \Big).
\eea
Then, adding both, the quadratic divergence is cancelled, so the SM mass parameter becomes for $f\gg v$
\bea
\Delta m^2_H\approx -\frac{N_c y^2_A}{4\pi^2}\,m^2_{t_B}\ln \Big(\frac{\Lambda}{m_{t_B}} \Big).
\eea

On the other hand, the one-loop quartic coupling for the SM Higgs is also obtained as follows,
\bea
\Delta\lambda_H= \frac{N_c}{8\pi^2}\, y^4_A \ln \Big(\frac{\Lambda}{m_{t_A}}\Big) + \frac{3N_c}{16\pi^2}\, y^4_B \ln \Big(\frac{\Lambda}{m_{t_B}}\Big). 
\eea
Then, for $y_A=y_B$, the electroweak VEV is given by
\bea
v=\sqrt{-\frac{m^2_H}{\lambda_H}}\sim  \frac{m_{t_B}}{y_A}\sim f. 
\eea
Therefore, in order to generate a mild hierarchy $v<f$ such that the cutoff scale is delayed to $\Lambda=4\pi f$ of order $5\,{\rm TeV}$,   we need to introduce a soft $Z_2$ breaking term by 
\bea
\Delta V= \mu^2 H^\dagger_A H_A.
\eea
As a result, from eq.~(\ref{HA-exp}), we get the additional corrections to both $m^2_H$ and $\lambda_H$ by
\bea
\Delta m^2_H &=& \mu^2,  \qquad
\Delta \lambda_H = - \frac{\mu^2}{3f^2}.
\eea
In this case, we can tune $\mu^2$ such that a correct electroweak symmetry breaking occurs at $v<f$.

\subsection{Four-form flux relaxation}

A recent proposal was made for relaxing the cosmological constant and the Higgs mass parameter to observed values by the same four-form flux \cite{membrane}.  A dimensionless coupling between the four-form flux and the Higgs field \cite{hierarchy,Giudice,Kaloper} was introduced such that the flux parameter is scannable in steps of weak-scale value to relax the Higgs mass parameter to a correct value without a fine-tuning. In this case, the scanning of the Higgs mass parameter stops at a right value for electroweak symmetry breaking as the tunneling probability from the dS phase just after the last membrane nucleation and the AdS phase is exponentially suppressed. 
There are other interesting ideas for connecting the Higgs mass problem to the cosmological constant problem \cite{others}. 

We introduce a three-index anti-symmetric tensor field $A_{\nu\rho\sigma}$, whose four-form field strength is given by  $F_{\mu\nu\rho\sigma}=4\, \partial_{[\mu} A_{\nu\rho\sigma]}$.
The Lagrangian including the four-form flux is given in the following,
\bea
{\cal L} = {\cal L}_0 +{\cal L}_{\rm ext} \label{full}
\eea
where 
\bea
 {\cal L}_0 =  \sqrt{-g} \bigg[\frac{1}{2}R  -\Lambda -\frac{1}{48} F_{\mu\nu\rho\sigma} F^{\mu\nu\rho\sigma}   |D_\mu H|^2 -M^2 |H|^2 +\lambda_H |H|^4 +  \frac{c_H}{24} \,\epsilon^{\mu\nu\rho\sigma} F_{\mu\nu\rho\sigma} \, |H|^2 \bigg] \label{L0}
\eea
and the extra Lagrangian ${\cal L}_{\rm ext}$ is composed of ${\cal L}_{\rm ext}={\cal L}_S+{\cal L}_L+ {\cal L}_{\rm memb} $ with
\bea
 {\cal L}_S &=&\frac{1}{6}\partial_\mu\bigg[\Big( \sqrt{-g}\,  F^{\mu\nu\rho\sigma} -c_H \epsilon^{\mu\nu\rho\sigma}  |H|^2  \Big)A_{\nu\rho\sigma} \bigg],  \\
 {\cal L}_L &=& \frac{q}{24}\, \epsilon^{\mu\nu\rho\sigma} \Big( F_{\mu\nu\rho\sigma}- 4\, \partial_{[\mu} A_{\nu\rho\sigma]} \Big),  \label{LL} \\
 {\cal L}_{\rm memb}&=& \frac{e}{6} \int d^3\xi\,  \delta^4(x-x(\xi))\, A_{\nu\rho\sigma} \frac{\partial x^\nu}{\partial \xi^a} \frac{\partial x^\rho}{\partial \xi^b} \frac{\partial x^\sigma}{\partial \xi^c} \,\epsilon^{abc} \nonumber \\
 &&-T\int d^3\xi\, \sqrt{-g^{(3)}}\,  \delta^4(x-x(\xi)).
\eea
Here, we note that ${\cal L}_S$ is the surface term for the well-defined variation of the action, $q$ is the Lagrange multiplier, which becomes the flux parameter due to the equation of motion, $e, T$ are the charge and tension of the membrane, respectively.   

Then, using  the equation of motion for $F_{\mu\nu\rho\sigma}$ \cite{hmlee1,hmlee2,hmlee3} as follows,
\bea
F^{\mu\nu\rho\sigma}=\frac{1}{\sqrt{-g}}\, \epsilon^{\mu\nu\rho\sigma} \Big(c_H |H|^2+q\Big),
\eea
and integrating out $F_{\mu\nu\rho\sigma}$, we recast the full Lagrangian (\ref{full}) into
\bea
{\cal L} &=&\sqrt{-g} \bigg[\frac{1}{2}R-\Lambda-  |D_\mu H|^2 +M^2 |H|^2 -\lambda_H |H|^4 -\frac{1}{2} (c_H |H|^2+q)^2 \bigg] +{\cal L}_{\rm nucl}\label{Lagfull}
\eea 
with
\bea
{\cal L}_{\rm nucl}= \frac{1}{6}\epsilon^{\mu\nu\rho\sigma} \partial_\mu q A_{\nu\rho\sigma} +\frac{e}{6} \int d^3\xi \, \delta^4(x-x(\xi))\, A_{\nu\rho\sigma} \frac{\partial x^\nu}{\partial \xi^a} \frac{\partial x^\rho}{\partial \xi^b} \frac{\partial x^\sigma}{\partial \xi^c} \epsilon^{abc}.  \label{extra}
\eea
Then, the effective Higgs mass parameter, the effective cosmological constant and the effective Higgs quartic coupling \cite{Giudice,Kaloper,hmlee1} are given by
\bea
M^2_{\rm eff}(q) &=& M^2 - c_H\, q,  \label{effHmass}\\
  \Lambda_{\rm eff} (q) &=& \Lambda + \frac{1}{2}\, q^2, \\
  \lambda_{H,{\rm eff}}&=&\lambda_H+\frac{1}{2}c^2_H
  \eea
Therefore, for $q>q_c$ with $q_c\equiv M^2/c_H$, the Higgs mass parameter in eq.~(\ref{effHmass}) becomes $M^2_{\rm eff}<0$, so electroweak symmetry is unbroken, whereas for $q<q_c$, we are in the broken phase for electroweak symmetry.  For $c_H={\cal O}(1)$ and the membrane charge $e$ of electroweak scale,  we obtain the observed Higgs mass parameter as $M^2_{\rm eff}= c_H\, e$, once the flux change stops at $q=q_c-e$ due to the suppression of a further tunneling with more membrane nucleation \cite{Giudice,Kaloper,hmlee1,hmlee2,hmlee3}.
For $\Lambda<0$, we can cancel a large cosmological constant by the contribution from the same flux parameter until $\Lambda_{\rm eff}$ takes the observed value at $q=q_c-e$, but we need to rely on an anthropic argument for that with $e$ being of order weak scale \cite{anthropic,Giudice}.  

The  four-form relaxation mechanism for the Higgs mass depends on the tunneling rate with membrane nucleation in the last stage of the four-form scanning. 
The tunneling rate from the last dS phase to the true vacuum depends on the bounce action $B$ for the instanton solution with radius ${\bar r}_0$ \cite{coleman,tunneling,hmlee1}, given in the following,
\bea
\gamma\equiv {\bar r}^{-4}_0\, e^{-B} \label{rate}
\eea
where the bounce action is given by
\bea
B=\frac{27\pi^2}{2} \, \frac{T^4}{(\Delta \Lambda)^3}\,\left( 1+\frac{1}{4} r^2_0 H^2\right)^{-2},
 \label{bounce}
\eea
with $r_0=\frac{3T}{\Delta \Lambda}$ being the instanton radius in the absence of gravity,
and the instanton radius ${\bar r}_0$ and the dS radius $H^{-1}$ are given, respectively, by 
\bea
{\bar r}_0=\frac{r_0}{1+\frac{1}{4} r^2_0 H^2}, \quad
H^{-1} = \frac{\sqrt{3} M_P}{\sqrt{\Delta\Lambda}}.
\eea
Here, $\Delta\Lambda\simeq e q_c$ is the change of the cosmological constant due to the last tunneling.
The gravitational corrections lead to suppressions for both the bounce action and the instanton radius. For $r_0< 2H^{-1}$, which corresponds to $\frac{T^2}{M^2_P}<\frac{4}{3} \Delta\Lambda$, we can ignore the curvature of the dS spacetime, so the tunneling rate becomes $\gamma\simeq r^{-4}_0\, e^{-B}$ with $B\simeq\frac{27\pi^2}{2} \, \frac{T^4}{(\Delta \Lambda)^3}$. On the other hand, for $r_0\gtrsim 2 H^{-1}$,  the bounce action is dominated by the curvature of the dS space, so the tunneling rate becomes $\gamma\simeq r^{-4}_0 \Big(\frac{r_0 H}{2} \Big)^8\, e^{-B}$ with $B\simeq \frac{24\pi^2 M^4_P}{\Delta\Lambda}$. 

We remark on the condition for the last dS phase at $q=q_c$ to become unstable within the Hubble volume, namely, $\gamma> H^4$.
In the case with $r_0< 2H^{-1}$ and $T=M^3_*$, we can obtain the condition on the brane tension for $\gamma> H^4$ \cite{hmlee1}, as follows,
\bea
M_*< \frac{1}{1.85^{1/12}}\, (\Delta \Lambda)^{1/4}\simeq \frac{1}{1.85^{1/12}}\, (eq_c)^{1/4}.  \label{instable}
\eea
Therefore, for $q_c\sim M^2_P$ and $e\sim (100\,{\rm GeV})^2$, the above instability bound becomes $M_*< 10^{10}\,{\rm GeV}$, being consistent with a negligible gravitational correction to the bounce action.  As a result, we would need an extra symmetry to keep the membrane tension to be much smaller than the Planck scale.

In the original scenario with the four-form flux, however, there is a need of reheating at the end of the membrane nucleation unless the non-perturbative particle production in the time-dependent background is efficient \cite{Giudice}. Otherwise, the Universe would be empty after the continuous exponential expansion in dS phases.
We give a schematic description of the reheating dynamics  in the following general form of the effective potential containing a singlet scalar field  $\phi$ \cite{hmlee2},
\bea
V(H,\phi) = V_{\rm eff}(H) + (k_1\phi^n+q+k_2)^2+ V_{\rm int}(\phi,H)
\eea
where $V_{\rm eff}(H)=-M^2_{\rm eff} |H|^2 +\lambda_{H,{\rm eff}} |H|^4$, and $k_1,k_2$ are constant parameters and $n$ is the positive integer, and $V_{\rm int}(\phi,H)$  is the interaction potential between the SM Higgs and the singlet scalar field.
Then, due to the flux-dependent minimum of the potential, the singlet scalar field can be displaced from the minimum after the last membrane nucleation such that the initial condition with a nonzero vacuum energy is realized. 

The maximum reheating temperature in this model can be inferred from the instantaneous reheating after the last membrane nucleation, 
\bea
\Delta \Lambda = \frac{\pi^2}{30}\,  g_*\, T^4_{\rm max},
\eea
that is,
\bea
T_{\rm max} = 8.5\times 10^9\,{\rm GeV} \, \bigg(\frac{100}{g_*}\bigg)^{1/4}\bigg(\frac{\sqrt{e}}{100\,{\rm GeV}} \bigg)^{1/2}\bigg(\frac{M^2}{c_H M^2_P}\bigg)^{1/4}.
\eea
But, the reheating temperature is model-dependent. Some concrete mechanisms for reheating in four-form flux models have been proposed, with the four-form couplings to gravity \cite{hmlee1,hmlee2} or a pseudo-scalar field $\phi$ \cite{hmlee2,hmlee4} or a complex scalar field $\Phi$ \cite{hmlee2}, in the following form,
\bea
{\cal L}_{{\rm RH},1}=-\frac{c_1}{24} \,\epsilon^{\mu\nu\rho\sigma} F_{\mu\nu\rho\sigma} \,R
\eea
or
\bea
{\cal L}_{{\rm RH},2}=\frac{\mu}{24} \,\epsilon^{\mu\nu\rho\sigma} F_{\mu\nu\rho\sigma} \, \phi
\eea
or
\bea
{\cal L}_{{\rm RH},3}=\frac{c_\Phi}{24} \,\epsilon^{\mu\nu\rho\sigma} F_{\mu\nu\rho\sigma}|\Phi|^2.
\eea
Here, $c_1, \mu, c_\Phi$ are extra four-form couplings in the reheating sector.
In all the above cases, the  four-form couplings give rise to the additional potential for a singlet scalar field with the flux-dependent minimum. The singlet scalar is responsible for reheating the Universe to a sufficiently high temperature as well as the production of dark matter \cite{hmlee2,hmlee4}.

\section{Dark matter physics}

There often appear new candidates for dark matter in some of solutions to the hierarchy problem, such as neutralinos in MSSM, mirror leptons in twin Higgs models \cite{twindm}, weakly coupled dark matter in clockwork models \cite{CWdm}, decaying dark matter in relaxion models \cite{relaxiondm}, etc. Strong bounds from direct detection can be easily evaded if dark matter annihilates into the SM singlet states or dark matter co-annihilate with a next-to-lightest particle. 

On the other hand, dark matter becomes naturally strongly coupled if it is a composite state due to a QCD-like dynamics as for mirror QCD with light quarks in twin Higgs models. Moreover, Wess-Zumino-Witten terms in a dark chiral perturbation theory \cite{WZW,SIMP-mesons,SIMP-mesons2,SIMP-mesons22, SIMP-mesons3} provide point-like $3\rightarrow 2$ processes for dark matter annihilation. 

There are a plenty of candidates for new dark gauge bosons beyond the SM, that could assist the annihilation of dark matter. There are hypercharge partners in twin Higgs models, and general light dark photons are present to guarantee the stability of dark matter. 

In this section, we give a brief overview on cosmology and thermodynamics (See Ref.~\cite{kolb}.).
Then, we summarize the detailed calculation of WIMP abundances. 
We also review some of new production mechanisms for light dark matter below GeV scale.

\subsection{Compact cosmology}

The Friedmann-Robertson-Walker metric,
\bea
ds^2= dt^2- a^2(t) \bigg[ \frac{dr^2}{1-k r^2}+ r^2d\theta^2+ r^2 \sin^2\theta d\phi^2  \bigg],
\eea 
describes the homogeneous and isotropic Universe, the dynamics of which is governed by the Friedmann equation,
\bea
H^2+\frac{k}{a^2} = \frac{\rho}{3M^2_{P}}, \qquad H= \frac{\dot a}{a}, \label{Fried}
\eea
where $k=+1,0,-1$ for closed, flat and open Universes and $\rho=\rho_M+\rho_R+\rho_\Lambda$,
and the continuity equation,
\bea
{\dot\rho}_i + 3H(\rho_i + p_i)=0.
\eea
The latter determines the energy density by $\rho_i\propto R^{-3(1+w_i)}$ with $w_i$ being the equation of state, $w_i=0, \frac{1}{3}, -1$, for matter, radiation and cosmological constant, respectively. The fractions of energy densities are defined by 
\bea
\Omega=\frac{\rho}{3M^2_P H^2}=\sum_i \Omega_i=1+ \frac{k}{(Ha)^2}.
\eea
From the Friedmann equation (\ref{Fried}), the scale factor $a(t)$  is obtained as $a(t)\propto t^{2/(3(w+1)}$.

The number density for particles with degrees of freedom $g$ is given by
\bea
n=\frac{g}{(2\pi)^3}\int d^3p  \, f({\vec p},t)
\eea
where $f({\vec p},t)$ is the occupancy distribution, taking
\bea
f({\vec p})= \frac{1}{e^{(E-\mu)/T}\pm 1}
\eea
for $+$ for Fermi-Dirac(FD) statistics and $-$ for Bose-Einstein(BE) statistics and $\mu$ is the chemical potential with $\mu=+1$ for FD and $\mu=-1$ for BE. 
The chemical equilibrium for $i+j\leftrightarrow k+l$ is achieved for $\mu_i+\mu_j=\mu_k+\mu_l$.  
For $f({\vec p},t)=f({\vec p})$, (ignoring the chemical potential), the number density for particles in equilibrium is
\bea
n_{\rm eq} =\bigg\{ \begin{array}{cc} \frac{g_{\rm eff} \zeta(3)}{\pi^2}\, T^3, \quad T\gtrsim m, \vspace{0.2cm} \\ 
g \Big(\frac{mT}{2\pi} \Big)^{3/2}\, e^{-m/T}, \quad T\lesssim m \end{array} 
\eea 
where $g_{\rm eff}$ is the number of effective degrees of freedom, taking $g_{\rm eff}= g$ for BS and $g_{\rm eff}= \frac{3}{4} g$ for FD, and $\zeta(3)=1.20206$. On the other hand, the energy density for particles in equilibrium is
\bea
\rho &=& \frac{g}{(2\pi)^3}\int d^3p  \, E\, f({\vec p}) \nonumber  \\
&=&\bigg\{ \begin{array}{cc} \frac{\pi^2}{30}\,  g_*\,T^4,  \quad T\gtrsim m,  \vspace{0.2cm}  \\ 
m \, n_{\rm eq}, \quad T\lesssim m \end{array} 
\eea
where $g_*$ is the number of effective massless degrees of freedom. 
Moreover, the entropy density is (for relativistic particles)
\bea
s=\frac{\rho+p}{T}= \frac{2\pi^2}{45}\, g_{*s} T^3
\eea
where $g_{*s}$ is the number of effective massless degrees of freedom in equilibrium.
The total entropy $S=a^3 s$ is conserved, so $g_{*s} T^3$=constant. 

During radiation domination, the Hubble parameter is given by
\bea
\frac{1}{2t} = H=\sqrt{\frac{\rho_R}{3M^2_P}} = 0.33 \, g^{1/2}_* \,\frac{T^2}{M_{Pl}}, 
\eea
resulting in the time to temperature conversion relation,
\bea
t= 1.515\,  g^{-1/2}_*\, \frac{M_{Pl}}{T^2}\sim \Big(\frac{T}{\rm MeV} \Big)^{-2}\,{\rm sec}. 
\eea

\subsection{WIMP}

Production mechanisms for Weakly Interacting Massive Particles (WIMP) \cite{WIMP} are reviewed.  
The general discussion is based on text books on cosmology such as Kolb and Turner \cite{kolb} and some of recent review articles \cite{dm0,dm1,dm2}.

\subsubsection{DM annihilations}

When $1+2\rightarrow 3+4$ processes change the number of dark matter particles, 
the Boltzmann equation determining the dark matter density is
\bea
\frac{dn}{dt} + 3H n = C_{2\rightarrow 2}
\eea
 with
 \bea
 C_{2\rightarrow 2}&=&-\frac{1}{s_i s_f}\int d\Pi_1 d\Pi_2 d\Pi_3 d\Pi_4 (2\pi)^4 \delta^4\Big(\sum p\Big) |M_{2\rightarrow 2}|^2 \bigg[ f_1 f_2(1\pm f_3)(1\pm f_4) \nonumber \\
 &&\quad  -f_3 f_4 (1\pm f_1)(1\pm f_2) \bigg] \nonumber \\
 &\simeq & -\frac{1}{s_i s_f}\int d\Pi_1 d\Pi_2 d\Pi_3 d\Pi_4 (2\pi)^4 \delta^4\Big(\sum p\Big) |M_{2\rightarrow 2}|^2 \Big(f_1 f_2 -f_3 f_4 \Big)
 \eea
where $d\Pi\equiv \frac{1}{(2\pi)^3} \frac{d^3 p}{ 2E}$, and $+$ for BS and $-$ for FD, and  $s_{i,f}=1(2)$ for two different (identical) particles in the initial or final states.  Here, we have ignored Pauli-blocking and stimulated emission factors so $1\pm f\simeq 1$.
In the non-relativistic limit for dark matter, 3 and 4 particles in equilibrium have $f_3=f^{\rm eq}_3=e^{-E_3/T}$ and $f_4=f^{\rm eq}_4=e^{-E_4/T}$. Then, from the energy conservation, $E_1+E_2=E_3+E_4$, we obtain $f_3 f_4 = e^{-(E_3+E_4)/T}=e^{-(E_1+E_2)/T}=f^{\rm eq}_1 f^{\rm eq}_2$. ($f^{\rm eq}_3 f^{\rm eq}_4=f^{\rm eq}_1 f^{\rm eq}_2$ is also true in general  and it is nothing but the detailed balance condition.) Moreover, for the slowly expanding Universe, we have $f_{1,2}\simeq f^{\rm eq}_{1,2}\, \Big(\frac{n_{1,2}}{n^{\rm eq}_{1,2}}\Big)$.
Then, for $n_{1}=n_2=n$ and $n^{\rm eq}_1=m^{\rm eq}_2=n_{\rm eq}$, the annihilation term becomes
\bea
 C_{2\rightarrow 2}\simeq  -\langle\sigma |v|\rangle (n^2- n^2_{\rm eq})
\eea
where the averaged annihilation cross section is
\bea
\langle\sigma |v|\rangle= \frac{1}{n^2_{\rm eq}}\, \frac{1}{s_i s_f} \int \frac{d^3p_1}{(2\pi)^3} \frac{ d^3p_2}{(2\pi)^3}\,  f^{\rm eq}_1 f^{\rm eq}_2 \, (\sigma|v|)
\eea
with
\bea
(\sigma|v|)\equiv \frac{1}{4E_1 E_2}\int d\Pi_3 d\Pi_4  (2\pi)^4 \delta^4\Big(\sum p\Big) |M_{2\rightarrow 2}|^2.
\eea
For comparison to the Hubble expansion rate $H$, the effective annihilation rate is defined as $\Gamma_{\rm ann}=n_{\rm eq} \langle\sigma |v|\rangle$.

\subsubsection{WIMP abundances}

We introduce $Y= \frac{n}{s}$ to estimate the total number of DM particles (or abundances) \cite{WIMP}.
Then, the Boltzmann equation becomes
\bea
{\dot n} + 3H \, n = s {\dot Y} =  -\langle\sigma |v|\rangle  s^2 (Y^2- Y^2_{\rm eq})
\eea
where 
\bea
Y_{\rm eq} &=& \frac{n_{\rm eq}}{s} \nonumber \\  
&=& \bigg\{\begin{array}{cc}0.278\, \frac{g_{\rm eff}}{g_{*s}}, \quad x\ll 3 \, ({\rm rel.}),  \vspace{0.2cm} \\ 
0.145\, \frac{g_{\rm eff}}{g_{*s}}\, x^{3/2}\, e^{-x},  \quad x\gg 3 \, ({\rm non-rel.}). \end{array} 
\eea
Then, using the variable $x=\frac{m}{T}$ and $\frac{d}{dt}=H(m)\, \frac{1}{x}\frac{d}{dx}$ with $H(m)=H(T=m)$, the Boltzmann equation becomes
\bea
\frac{dY}{dx} = -\frac{x\langle\sigma |v|\rangle s }{H(m)}\, (Y^2- Y^2_{\rm eq}) = -\frac{n_{\rm eq}\langle\sigma |v|\rangle }{xH Y_{\rm eq}}\, (Y^2- Y^2_{\rm eq}).
\eea
That is,
\bea
\frac{x}{Y_{\rm eq}}\, \frac{dY}{dx} =  -\frac{\Gamma_{\rm ann}}{H}\, \bigg[ \bigg( \frac{Y}{Y_{\rm eq}}\bigg)^2-1\bigg], \qquad \Gamma_{\rm ann}=n \langle\sigma |v|\rangle.
\eea
Therefore, the DM abundance freezes out as $-\frac{\Delta Y}{Y}\sim \frac{x}{Y_{\rm eq}}\, \frac{dY}{dx} \sim -\frac{\Gamma_{\rm ann}}{H}\lesssim 1$, so it is determined at $x=x_f$ at which $\Gamma_{\rm ann}\sim H$.  

Suppose that dark matter freezes out when non-relativistic. Then, for $\langle\sigma |v|\rangle =\sigma_0\, x^{-n}$, we rewrite the Boltzmann equation as
\bea
\frac{dY}{dx} = -\lambda x^{-n-2} \, (Y^2- Y^2_{\rm eq}) \label{Yeq}
\eea
with
\bea
\lambda\equiv \bigg[ \frac{\langle\sigma |v|\rangle s }{H(m)}\bigg]_{x=1} =1.329\,  (g_{*s}/g^{1/2}_*) M_{Pl} m \sigma_0. 
\eea

The master formula for dark matter abundance at present is 
\bea
\Omega_{\rm DM} h^2 = \frac{\rho_{{\rm DM}}}{3M^2_P H^2_0/h^2}= 0.2745\, \Big(\frac{Y_\infty}{10^{-11}}\Big)\, \Big(\frac{m}{100\,{\rm GeV}} \Big) \label{relic}
\eea
where $\rho_{{\rm DM}}=m\, Y_\infty$ with $Y_\infty=Y(T=0)$ is used.

For $Y\gg Y_{\rm eq}$, the Boltzmann equation (\ref{Yeq}) becomes
\bea
\frac{dY}{dx} \simeq  -\lambda x^{-n-2} \, Y^2.
\eea
We get  $Y_\infty\equiv Y(x=\infty)$ as
\bea
\frac{1}{Y_\infty} \simeq \frac{1}{Y(x_f)}+\frac{\lambda}{n+1}\, x^{-n-1}_f, 
\eea
resulting in
\bea
Y_\infty\simeq \frac{n+1}{\lambda}\, x^{n+1}_f= \frac{0.75(n+1) g^{1/2}_*}{g_{*s}}\, \frac{x^{n+1}_f}{M_{Pl} m\,\sigma_0}.
\eea
So, the DM abundance is inversely proportional to the annihilation cross section, $\sigma_0$.

From the freeze-out condition $\Gamma_{\rm ann}=n_{\rm eq}\langle\sigma |v|\rangle\simeq H$, we get the freeze-out temperature, $T_f=m/x_f$, with
\bea
x_f&=&\ln \Big(0.038 c(c+2)g/g^{1/2}_* M_{Pl} m\sigma_0  \Big) \nonumber \\
&&-\Big(n+\frac{1}{2} \Big) \ln \Big[\ln \Big(0.038 c(c+2)g/g^{1/2}_* M_{Pl} m\sigma_0   \Big) \Big]
\eea
where $Y(x_f)=(c+1)Y_{\rm eq}(x_f)$ with $c={\cal O}(1)$. 
So, for $\sigma_0=\frac{\alpha^2}{m^2}$, $g=2$ and $n=0$ (s-wave annihilation), we get
\bea
x_f\simeq 20 -\frac{1}{2} \ln\Big( \frac{g_*}{61.75}\Big) + 2 \ln\Big(\frac{\alpha}{1/30} \Big) -\ln \Big( \frac{m}{600\,{\rm GeV}}\Big).
\eea
Then, from eq.~(\ref{relic}) with $g_{*s}=g_*$, the WIMP relic abundance is
\bea
\Omega_{\rm DM} h^2 = 0.1 \Big(\frac{61.75}{g_*} \Big)^{1/2} \Big(\frac{x_f}{20}\Big) \Big(\frac{1/30}{\alpha}\Big)^2 \Big( \frac{m}{600\,{\rm GeV}}\Big)^2.
\eea

\subsection{Production mechanisms for light dark matter}

WIMP dark matter is strongly constrained by direct detection experiments.
On the other hand, light dark matter has drawn a lot of attention in view of new detection strategies \cite{LDM1,LDM2} and self-interactions of dark matter can provide solutions to the small-scale problems in galaxy scales \cite{SIDM}. 
But, the thermal cross section of light dark matter below $50\,{\rm GeV}$ is ruled out by CMB at recombination \cite{planck}. Thus, there is a need to develop new production mechanisms for light dark matter.   
There are co-annihilation \cite{except}, DM self-interactions \cite{SIMP,SIMP-mesons,SIMPm,SIMP2,SIMP-mesons2,SIMP-mesons22,SIMP-mesons3,elder,cannibal}, forbidden channels \cite{FDM}, co-scattering \cite{coscatt}, co-decay \cite{codecay}, etc, proposed as new production mechanisms for light dark matter.  We focus on DM self-interactions and forbidden channels and discuss the kinetic equilibrium condition that is crucial for light dark matter.

\subsubsection{DM production from self-interactions}

Strongly Interacting Massive Particles (SIMPs) have the abundance determined by $3\rightarrow 2$ processes with  large self-interactions, instead of annihilating in pairs \cite{SIMP}. 
 
Assuming that $2\rightarrow 2$ processes are subdominant,  we have  the Boltzmann equation for the relic density with $1+2+3\rightarrow 4+5$ as
\bea
\frac{dn}{dt} + 3H n = C_{3\rightarrow 2}
\eea
 with
 \bea
 C_{3\rightarrow 2}&=&-\frac{1}{s_i s_f}\int d\Pi_1 d\Pi_2 d\Pi_3 d\Pi_4d\Pi_5 (2\pi)^4 \delta^4\Big(\sum p\Big) |M_{3\rightarrow 2}|^2 \bigg[ f_1 f_2f_3 (1\pm f_4)(1\pm f_5) \nonumber \\
 &&\quad  -f_4 f_5 (1\pm f_1)(1\pm f_2)(1\pm f_3) \bigg] \nonumber \\
 &\simeq & -\frac{1}{s_i s_f}\int d\Pi_1 d\Pi_2 d\Pi_3 d\Pi_4d\Pi_5 (2\pi)^4 \delta^4\Big(\sum p\Big) |M_{3\rightarrow 2}|^2 \Big(f_1 f_2f_3 -f_4 f_5 \Big)
 \eea
where $s_i=n_i!$ for $n_i$ identical particles in the initial states and  $s_f=n_f!$ for $n_f$ identical particles in the final states. 
Using $f_{i}\simeq f^{\rm eq}_{i}\, \Big(\frac{n_{i}}{n^{\rm eq}_{i}}\Big)$ and $E_1+E_2+E_3=E_4+E_5$, 
we obtain the annihilation term,
\bea
 C_{3\rightarrow 2}&=&-\frac{1}{n^{\rm eq}_1 n^{\rm eq}_2 n^{\rm eq}_3} \frac{1}{s_i s_f}\int d\Pi_1 d\Pi_2 d\Pi_3 d\Pi_4d\Pi_5 (2\pi)^4 \delta^4\Big(\sum p\Big) |M_{3\rightarrow 2}|^2  f^{\rm eq}_1 f^{\rm eq}_2 f^{\rm eq}_3  \nonumber \\
 &&\quad \times \bigg(n_1 n_2 n_3 - \frac{n^{\rm eq}_1 n^{\rm eq}_2 n^{\rm eq}_3  }{ n^{\rm eq}_4n^{\rm eq}_5  }\, n_4 n_5 \bigg). 
\eea
For $n_1=n_2=n_3=n$ and $n^{\rm eq}_1=n^{\rm eq}_2=n^{\rm eq}_3=n_{\rm eq}$ , the above result becomes simplified to
\bea
C_{3\rightarrow 2}=-\langle\sigma v^2\rangle  (n^3- n_{\rm eq} n^2)
\eea
where the averaged annihilation ``cross section'' is
\bea
\langle\sigma v^2\rangle =\frac{1}{n^3_{\rm eq}} \,  \frac{1}{s_i s_f} \int \frac{d^3p_1}{(2\pi)^3} \frac{ d^3p_2}{(2\pi)^3}\frac{ d^3p_3}{(2\pi)^3}\,  f^{\rm eq}_1 f^{\rm eq}_2f^{\rm eq}_3 \, (\sigma v^2)
\eea
with
\bea
(\sigma v^2)\equiv \frac{1}{8E_1 E_2 E_3} \int d\Pi_3 d\Pi_4  (2\pi)^4 \delta^4\Big(\sum p\Big) |M_{3\rightarrow 2}|^2.
\eea
Here, $[M_{3\rightarrow 2}]=E^{-1}$ so  $[(\sigma v^2)]=E^{-5}$, as compared to  the standard cross section, $[(\sigma|v|)]=E^{-2}$.
We note that the effective $2\rightarrow 2$ annihilation rate is $\Gamma_{\rm ann}=n^2_{\rm eq} \langle\sigma v^2\rangle$, as compared to the $2\rightarrow 2$ annihilation rate, $\Gamma_{\rm ann}=n_{\rm eq}\langle\sigma |v|\rangle$.
The thermal average of the $3\rightarrow 2$ cross section needs caution in the case of velocity-dependence or resonance poles \cite{SIMP2,resonances}.

Similarly to the WIMP case,  after a change of variable to $x=m/T$, we can rewrite the Boltzmann equation as
\bea
\frac{dY}{dx} = -\frac{x\langle\sigma v^2\rangle s^2 }{H(m)}\, (Y^3- Y^2 Y_{\rm eq}) = -\frac{n^2_{\rm eq}\langle\sigma v^2\rangle }{xH Y^2_{\rm eq}}\, (Y^3- Y^2 Y_{\rm eq}).
\eea
That is,
\bea
\frac{x}{Y_{\rm eq}}\, \frac{dY}{dx} =  -\frac{\Gamma_{\rm ann}}{H}\, \bigg[ \bigg( \frac{Y}{Y_{\rm eq}}\bigg)^3- \bigg(\frac{Y}{Y_{\rm eq}}\bigg)^2\bigg], \qquad \Gamma_{\rm ann}=n^2_{\rm eq} \langle\sigma v^2\rangle.
\eea
Therefore, the DM abundance freezes out as $-\frac{\Delta Y}{Y}\sim \frac{x}{Y_{\rm eq}}\, \frac{dY}{dx} \sim -\frac{\Gamma_{\rm ann}}{H}\lesssim 1$, so it is determined at $x=x_f$ at which $\Gamma_{\rm ann}\sim H$.  

Suppose that SIMP dark matter freezes out when non-relativistic. Then, for $\langle\sigma v^2\rangle =\sigma_0\, x^{-n}$, we rewrite the Boltzmann equation as
\bea
\frac{dY}{dx} = -\kappa x^{-n-5} \, (Y^3- Y^2 Y_{\rm eq}) \label{Yeq2}
\eea
with
\bea
\kappa\equiv \bigg[ \frac{\langle\sigma |v|\rangle s^2 }{H(m)}\bigg]_{x=1} =0.583\,  (g^2_{*s}/g^{1/2}_*) M_{Pl} m^4 \sigma_0. 
\eea

For $Y\gg Y_{\rm eq}$, the Boltzmann equation (\ref{Yeq2}) becomes
\bea
\frac{dY}{dx} \simeq  -\kappa x^{-n-5} \, Y^3.
\eea
We get  $Y_\infty\equiv Y(x=\infty)$ as
\bea
Y_\infty\simeq \sqrt{\frac{n+4}{2\kappa}}\, x^{(n+4)/2}_f= \frac{1.85\,(n/2+2)^{1/2} g^{1/4}_*}{g_{*s}}\, \frac{x^{(n+4)/2}_f}{(M_{Pl} m^4\,\sigma_0)^{1/2}}.
\eea
So, the DM abundance is inversely proportional to $(\sigma_0)^{1/2}$.

From the freeze-out condition $\Gamma_{\rm ann}=n_{\rm eq}^2\langle \sigma v^2\rangle\simeq H$, we also get the freeze-out temperature, $T_f=m/x_f$, with
\bea
x_f&=&\ln \Big(0.110 c(c+1)^2 g/g^{1/4}_* (M_{Pl} m^4\sigma_0)^{1/2}  \Big) \nonumber \\
&&-\frac{1}{2} (n+2) \ln \Big[\ln \Big(0.110 c(c+1)^2g/g^{1/4}_* (M_{Pl} m^4\sigma_0 )^{1/2}  \Big) \Big]
\eea
where $Y(x_f)=(c+1)Y_{\rm eq}(x_f)$ with $c={\cal O}(1)$. 
So, for $\sigma_0=\frac{\alpha^3}{m^5}$, $g=1$ and $n=0$ (s-wave annihilation), we get
\bea
x_f\simeq 18 -\frac{1}{4} \ln\Big( \frac{g_*}{10.75}\Big) + \frac{3}{2} \ln(\alpha) -\frac{1}{2}\ln \Big( \frac{m}{1\,{\rm GeV}}\Big).
\eea
Then, from eq.~(\ref{relic}) with $g_{*s}=g_*$, the WIMP relic abundance is
\bea
\Omega_{\rm DM} h^2 = 0.1 \bigg(\frac{10.75}{g_*} \bigg)^{3/4} \Big(\frac{x_f}{20}\Big)^2 \Big(\frac{4}{\alpha}\Big)^{3/2} \Big( \frac{m}{100\,{\rm MeV}}\Big)^{3/2}.
\eea

{\bf Problem}: Suppose a real scalar dark matter $\phi$ with mass $m$ and the interaction Lagrangian ${\cal L}_{\rm int}=-\frac{1}{3!}\,\kappa\, \phi^3$. Then, compute $|M_{3\rightarrow 2}|^2$ for $\phi\phi\phi\rightarrow \phi\phi$ and show the parameter space for $m$ and $\kappa$ satisfying the correct relic abundance, and compare it with the bound on the self-scattering cross section, $\sigma_{\rm scatt}/m<1\,{\rm cm^2/g}$ (Bullet cluster bound).

\subsubsection{Kinematics for SIMP dark matter}

We consider the kinematics of $3\rightarrow 2$ processes, $1+2+3\rightarrow 4+5$, and derive the formula for the corresponding cross section.  We take the most general masses for particles. 

In the center of mass frame for $4, 5$ particles, ${\vec p}_4+{\vec p}_5={\vec p}_1+{\vec p}_2+{\vec p}_3$, we have
\bea
\sqrt{s}=E_4 + E_5 = \sqrt{m^2_4+{\vec p}^2_{4,{\rm CM}}}=\sqrt{m^2_5+{\vec p}^2_{4,{\rm CM}}}.
\eea
Then, taking the square and solving for ${\vec p}_4$, we get
\bea
|{\vec p}_{4, {\rm CM}}| = \frac{1}{2} \sqrt{s}\, \sqrt{1-\frac{(m_4-m_5)^2}{s}}\, \sqrt{1-\frac{(m_4+m_5)^2}{s}}.
\eea
For $m_1=m_2=m_3=m_4=m_5=m$, it becomes 
\bea
|{\vec p}_{4,{\rm CM}}| = \frac{1}{2} \sqrt{s}\,  \sqrt{1-\frac{4m^2}{s}}\simeq \frac{\sqrt{5}}{2}\,m
\eea
in the nonrelativistic limit. 

The cross section for $3\rightarrow 2$ processes is
\bea
(\sigma v^2)&=& \frac{1}{128\pi^2E_1 E_2 E_3} \int \frac{|{\vec p}_4|^2 d|{\vec p}_4| d\Omega}{E_4 E_5}  \, \delta(E_1+E_2+E_3-E_4-E_5)|M_{3\rightarrow 2}|^2 \nonumber \\
&=& \frac{1}{128\pi^2E_1 E_2 E_3} \int d\Omega\, \frac{|{\vec p}_{4,{\rm CM}}|}{\sqrt{s}}\, |M_{3\rightarrow 2}|^2.  
\eea
Here, we have used 
\bea
\delta(E_1+E_2+E_3-E_4-E_5)=\delta(\sqrt{s}-E_4-E_5)=\bigg(\frac{|{\vec p}_4|}{E_4} +\frac{|{\vec p}_4|}{E_5}\bigg)^{-1} \delta({\vec p}_4-{\vec p}_{4, {\rm CM}}).
\eea

Then, for angle-independent $ |M_{3\rightarrow 2}|^2$ and $E_1\simeq m_1$, $E_2\simeq m_2$ and $E_3\simeq m_3$, we obtain
\bea
(\sigma v^2)&=& \frac{1}{32\pi E_1 E_2 E_3}\,  \frac{|{\vec p}_{4,{\rm CM}}|}{\sqrt{s}}\, |M_{3\rightarrow 2}|^2 \nonumber \\
&\simeq &    \frac{|M_{3\rightarrow 2}|^2}{64\pi E_1 E_2 E_3}\,   \sqrt{1-\frac{(m_4-m_5)^2}{(m_1+m_2+m_3)^2}}\, \sqrt{1-\frac{(m_4+m_5)^2}{(m_1+m_2+m_3)^2}}.
\eea
For  $m_1=m_2=m_3=m_4=m_5=m$, it becomes
\bea
(\sigma v^2)\simeq \frac{\sqrt{5}}{192\pi m^3}\, |M_{3\rightarrow 2}|^2. 
\eea

We can compare with  the formula for $2\rightarrow 2$ processes,
\bea
(\sigma |v|) = \frac{1}{64\pi^2E_1 E_2} \int d\Omega\, \frac{|{\vec p}_{4,{\rm CM}}|}{\sqrt{s}}\, |M_{2\rightarrow 2}|^2=  \frac{1}{16\pi E_1 E_2} \int d\Omega\, \frac{|{\vec p}_{4,{\rm CM}}|}{\sqrt{s}}\, |M_{2\rightarrow 2}|^2. 
\eea
In this case, for $m_1=m_2=m$ and ignoring the masses of the final states, the $2\rightarrow 2$ annihilation cross section in the non-relativistic limit is
\bea
(\sigma_{\rm ann} |v|)\simeq \frac{\sqrt{5}}{32\pi m^2}\, |M_{\rm ann}|^2. 
\eea
Similarly, the cross section for self-scattering cross section is 
\bea
\sigma_{\rm scatt} = \frac{1}{64\pi m^2}\, |M_{\rm scatt}|^2. 
\eea

\subsubsection{Kinetic equilibrium}

The temperature of dark matter is defined as
\bea
T_{\rm DM}= \frac{2}{3} \bigg\langle \frac{p^2}{2m}\bigg\rangle = \frac{1}{3} m\langle v^2\rangle.
\eea
If dark matter is in kinetic equilibrium with the SM plasma, dark matter temperature is the same as photon temperature. Otherwise, dark matter  temperature  evolves in time differently from photon temperature.

In the presence of the annihilation and elastic scattering of dark matter, the change of kinetic energy (or transfer of excess kinetic energy into the SM plasma) is given \cite{kineq2} by
\bea
{\dot K} &=& K_{\rm ann}\, \frac{\dot n}{n} + T\gamma(T) \nonumber \\
&\simeq & - K_{\rm ann}\,  m H T^{-1} + T\gamma(T)
\eea
where $ \frac{\dot n}{n}\simeq  \frac{{\dot n}_{\rm eq}}{n_{\rm eq}}\simeq -\frac{m}{T}\, H$, and  $K_{\rm ann}$ is the kinetic energy released per DM annihilation, given by $K_{\rm ann}=T$ for WIMP and $K_{\rm ann}=m$ for SIMP, and $\gamma(T)$ is the momentum relaxation rate \cite{kineq1,kineq2} for DM-$f_i$ elastic scattering with $g_i$ being the number of degrees of freedom,
\bea
\gamma(T) = \frac{g_i}{6m T}\, \int \frac{d^3 p}{(2\pi)^3}\, f_i (1\pm f_i)\, \frac{|{\vec p}_i|}{E_i}\, \sigma_{T,i} 
\eea
with
\bea
\sigma_{T,i} = \int^0_{-4 p^2} dt \, (-t) \,\frac{d\sigma_i}{dt}, \qquad \frac{d\sigma_i}{dt}= \frac{1}{64\pi m^2 k^2} \, \overline{|M_{Xf_i\rightarrow Xf_i}|^2}.
\eea
The kinetic equilibrium condition is 
\bea
\gamma(T)= \frac{m K_{\rm ann}}{T^2}\, H=\left\{\begin{array}{cc}  \Big(\frac{m}{T}\Big)\, H,\quad {\rm WIMP}  \vspace{0.2cm} \\   \Big(\frac{m}{T}\Big)^2\, H, \quad {\rm SIMP}. \end{array} \right.
\eea
We note that the transfer of kinetic energy per  DM scattering is $K_{\rm scatt}=\frac{q^2}{2m}\sim \frac{T^2}{m}$, so the number of scatterings to absorb the DM kinetic energy is $K_{\rm ann}/K_{\rm scatt}\sim m/T$ for WIMP and $\sim (m/T)^2$ for SIMP.  Thus, we need $m/T$ times more scatterings for SIMP than for WIMP.  We have an approximate formula for $T\gamma(T)\sim \langle n_{\rm SM}\sigma_{\rm scatt} |v| \,K_{\rm scatt}\rangle\sim \langle n_{\rm SM}\sigma_{\rm scatt} |v| \,T^2/m\rangle$.

{\bf Problem}: Compute the momentum relaxation rate for a complex scalar dark matter $\phi$ with mass $m$ that is charged under a dark $U(1)$. Here, we assume that the dark gauge boson $X_\mu$ has mass $m_X$ and has a kinetic mixing with the SM hypercharge gauge boson by ${\cal L}_{\rm mix}=-\frac{1}{2}\,\xi B_{\mu\nu} X^{\mu\nu}$ where $X_{\mu\nu}=\partial_\mu X_\nu-\partial_\nu X_\mu$.

\subsubsection{Forbidden dark matter}

Forbidden dark matter relies on $2\rightarrow 2$ annihilation processes that are kinematically forbidden at zero temperature but available at high temperature \cite{except,FDM}. 
Then, forbidden channels can determine the relic density in the early Universe, provided that the corresponding $2\rightarrow 2$ cross section is large enough.

When dark matter $\chi$ is lighter than a hidden sector particle, such as dark gauge boson $Z'$, with mass $m_{Z'}>m_\chi$, the Boltzmann equation for $n_\chi=n_{\chi^*}=n/2$ containing the forbidden channels is 
 \bea
\frac{d n}{dt} + 3H n =-\frac{1}{2}\langle \sigma |v|\rangle_{\chi\chi^*\rightarrow Z'Z'} n^2+2\langle\sigma |v|\rangle_{Z'Z'\rightarrow \chi\chi^*} (n^{\rm eq}_{Z'})^2.
 \eea
The detailed balance conditions at high temperature leads to the cross section for forbidden channels,
\bea
\langle\sigma |v|\rangle_{\chi\chi^*\rightarrow Z'Z'} &=&\frac{4 (n^{\rm eq}_{Z'})^2}{(n_{\rm eq})^2}\langle\sigma |v|\rangle_{Z'Z'\rightarrow \chi\chi^*} \nonumber \\
&=& 9 (1+\Delta)^3 e^{-2\Delta  x}\, \langle\sigma v\rangle_{Z'Z'\rightarrow \chi\chi^*} \label{Zp1}
\eea
with $\Delta \equiv (m_{Z'}-m_\chi)/m_\chi$.

We can rewrite the Boltzmann equation with the detailed balance conditions, (\ref{Zp1}),  as follows,
\bea
\frac{dY}{dx}=-\zeta x^{-2}   \left(\frac{9}{2}(1+\Delta_{Z'})^3 e^{-2\Delta x} \,Y^2-2(Y^{\rm eq}_{Z'})^2
\right)
\label{Zp-Boltz}
\eea
with
\bea
\zeta \equiv \frac{s(m_\chi)}{H(m_\chi)}\, \langle\sigma |v|\rangle_{Z'Z'\rightarrow \chi\chi^*}.
\eea
Then, when $Z'Z'\rightarrow \chi\chi^*$  is  s-wave,
 the approximate solution to the Boltzmann equation (\ref{Zp-Boltz}) is given by
\be
Y_\infty\approx \frac{x_f}{\zeta}\, e^{2\Delta x_f}\, g(\Delta_{Z'}, x_f)
\ee
with
\bea
g(\Delta,x_f)&=& \bigg[\frac{9}{2}\,(1+\Delta)^3  \Big(1-2(\Delta x_f)\, e^{2\Delta x_f}  \int^\infty_{2\Delta x_f} dt \,t^{-1} e^{-t} \Big) \bigg]^{-1} .
\eea
Consequently, the relic density is determined to be
\bea
\Omega_{\rm DM} h^2 
=0.2 \,  \Big(\frac{g_*}{10.75} \Big)^{-1/2}\Big(\frac{x_f}{20}\Big)\,g(\Delta, x_f)\, \bigg(\frac{ 1\,{\rm pb}}{e^{2\Delta x_f}\langle\sigma |v|\rangle_{Z'Z'\rightarrow \chi\chi^*}}\bigg).
\eea
Then, the $2\rightarrow 2$ annihilation cross section can be large, due to the inverse of the Boltzmann suppression factor, $e^{2\Delta x_f}$, being compatible with the relic density. 
In this case, the self-scattering cross section for $\chi\chi^*\rightarrow \chi\chi^*$  can be also large.

\section{Conclusions}

We have given an overview on the theoretical problems in the SM and the basics of supersymmetry suggested as a solution to the hierarchy problem. We have also touched upon some key points of new recent proposals for the hierarchy problem and discussed the production mechanisms for thermal dark matter such as WIMP, SIMP, etc. Interestingly, new dynamical degrees of freedom in each of the proposals play the role of a bridge in connecting to dark matter physics, being testable at the future collider and cosmology frontiers.

\section*{Acknowledgments}

The author appreciate deeply discussion and collaboration with many colleagues on interesting topics in particles physics and cosmology.
The work is supported in part by Basic Science Research Program through the National Research Foundation of Korea (NRF) funded by the Ministry of Education, Science and Technology (NRF-2019R1A2C2003738 and NRF-2021R1A4A2001897).

\end{document}